\documentclass[preprint,a4paper,notoc]{JHEP3}
\usepackage{amsmath}
\usepackage{epsfig}
\usepackage{multicol}
\usepackage{dsfont}
\usepackage{psfrag}
\usepackage{amssymb}
\usepackage{bbm}
\usepackage{cite}

\title{Non-perturbative renormalization of static-light four-fermion operators in quenched lattice QCD}

\author{\epsfig{figure=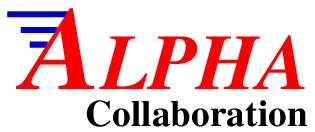,width=2.5cm}}

\author{
  Filippo Palombi\\
  DESY, Platanenallee 6, D-15738 Zeuthen, Germany\\
  E-mail: \email{filippo.palombi@desy.de}}
\author{
  Mauro Papinutto\\
  CERN, Physics Department, Theory Division, CH-1211 Geneva 23,
  Switzerland\\
  E-mail: \email{mauro.papinutto@cern.ch}}
\author{
  Carlos Pena\\
  CERN, Physics Department, Theory Division, CH-1211 Geneva 23,
  Switzerland\\
  E-mail: \email{carlos.pena.ruano@cern.ch}}
\author{
  Hartmut Wittig\\
  Institut f\"ur Kernphysik, University of Mainz, D-55099 Mainz, Germany\\
  E-mail: \email{wittig@kph.uni-mainz.de}}

\preprint{CERN-PH-TH/2007-097\\ DESY 07-090\\ MKPH-T-0709 \\ June 2007}

\abstract{We perform a non-perturbative study of the scale-dependent
renormalization factors of a multiplicatively renormalizable basis of
$\Delta{B}=2$ parity-odd four-fermion operators in quenched lattice
QCD.  Heavy quarks are treated in the static approximation with
various lattice discretizations of the static action. Light quarks are
described by non-perturbatively ${\rm O}(a)$ improved Wilson-type
fermions.  The renormalization group running is computed for a family
of Schr\"odinger functional (SF) schemes through finite volume
techniques in the continuum limit. We compute non-perturbatively the
relation between the renormalization group invariant operators and
their counterparts renormalized in the SF at a low energy scale.
Furthermore, we provide non-perturbative estimates for the matching
between the lattice regularized theory and all the SF schemes considered.}


\usepackage[OT2,OT1]{fontenc}
\newcommand\cyr{%
\renewcommand\rmdefault{wncyr}%
\renewcommand\sfdefault{wncyss}%
\renewcommand\encodingdefault{OT2}%
\normalfont
\selectfont}
\DeclareTextFontCommand{\textcyr}{\cyr} 

\newcommand{\parbreak}{\vskip 0.20cm}
\newcommand{\rR}{{\rm\scriptscriptstyle R}}
\newcommand{\RGI}{{\rm\scriptscriptstyle RGI}}

\newcommand{\cO}{{\cal O}}
\newcommand{\cQ}{{\cal Q}}

\newcommand{\cS}{{\cal S}}

\newcommand{\cZ}{{\cal Z}}

\newcommand{\Nf}{N_{\rm\scriptsize f}}

\newcommand{\Oa}{\mbox{O}(a)}

\newcommand{\Dslash}{\relax{\kern+.25em / \kern-.70em D}}
\newcommand{\lp}{{\scriptscriptstyle +}}

\newcommand{\lpm}{{\scriptscriptstyle \pm}}

\newcommand{\hopc}{\kappa_{\rm cr}}
\newcommand{\gbar}{\bar{g}}

\newcommand{\OVApAV}{\cO^\lp_{\rm\scriptscriptstyle VA+AV}}
\newcommand{\OSPpPS}{\cO^\lp_{\rm\scriptscriptstyle PS+SP}}
\newcommand{\OVAmAV}{\cO^\lp_{\rm\scriptscriptstyle VA-AV}}
\newcommand{\OSPmPS}{\cO^\lp_{\rm\scriptscriptstyle SP-PS}}

\begin{document}

\section{Introduction}

Particle-antiparticle transformations of neutral $B$-mesons are
currently being investigated in the framework of major experimental
programmes, aiming to constrain the top-quark sector of the
Cabibbo-Kobayashi-Maskawa (CKM) matrix. Measurements of the
oscillation frequencies $\Delta m_q$ $(q=d,s)$ allow to extract
$|V_{tq}|$, once the quantum mechanical amplitudes responsible for the
elementary transitions are known. The latter are customarily
represented in terms of the $B$-parameters $B_{B_q}$ through an
explicit factorization of the vacuum-saturation contribution, namely
\begin{equation}
\label{Bpars}
 \langle \bar B^0_q|
 \cO^{\scriptscriptstyle\Delta B=2}_{\rm\scriptscriptstyle LL}
 |B^0_q\rangle = \frac{8}{3}B_{B_q} f_{B_q}^2m_{B_q}^2\ .
\end{equation} 
A theoretical computation of the matrix element in Eq.~(\ref{Bpars})
requires non-perturbative techniques for a proper description of the
low-energy dynamics of the $B$-mesons. Lattice QCD is the obvious
methodology, insofar as all its systematic uncertainties can be
reduced to an acceptable level. We refer the reader to
\cite{Okamoto:2005zg,Onogi:2006km} for a review of recent lattice
determinations of the mixing parameters of the $B$-mesons.

\parbreak In a previous work \cite{Palombi:2006pu}, we have devised a
novel strategy to compute the above matrix elements in lattice QCD,
based on Heavy Quark Effective Theory (HQET) at leading order in the
heavy quark mass expansion in conjunction with twisted mass QCD
(tmQCD) \cite{Frezzotti:2000nk} in the light quark sector. This has
the advantage of removing the unwanted mixings under renormalization,
which arise with ordinary Wilson-type lattice fermions. The main idea
behind the proposed approach originates from the general observation
that in presence of a chirality breaking lattice regularization, like
the Wilson one, parity-odd four-fermion operators can have simpler
renormalization properties than their parity-even counterparts, as
also shown in \cite{Donini:1999sf}. In the particular case where light
quarks are described as Wilson fermions and heavy quarks are treated
in the static approximation, it is even possible to define a complete
basis $\left\{{\cQ'}_k^\lpm\right\}_{k=1,\dots,4}$ of multiplicatively
renormalizable parity-odd $\Delta B=2$ four-fermion operators, which
is given in sect.~\ref{sec_operators} below (see also Eq.~(2.12) of
\cite{Palombi:2006pu}). As derived in \cite{Palombi:2006pu}, this
result is mainly due to the heavy quark spin symmetry and time reversal, 
which strongly constrain the chirality breaking pattern, especially in the
parity-odd sector. The adoption of tmQCD allows to take advantage of such
properties by relating the parity-even operators of the effective
static theory entering the computation of $B_{B_q}$ to the
aforementioned operator basis, viz. ${\cQ'}_1^\lp$ and ${\cQ'}_2^\lp$.
Accordingly, the additional mixing under renormalization
with Wilson-type lattice fermions is avoided, thus
opening the way to a determination of the $B$-parameters with
reduced systematic uncertainties.

\parbreak This paper is devoted to a non-perturbative study of the
renormalization group (RG) running of the operator basis
$\left\{{\cQ'}_k^\lp\right\}_{k=1,\dots,4}$ in the quenched
approximation. Renormalization constants are defined in terms of
Schr\"odinger functional (SF) correlators, where periodic boundary
conditions (up to a phase $\theta$ for light-quark fields) are imposed 
along the spatial directions and Dirichlet boundary conditions are 
imposed in time. The SF formalism \cite{Luscher:1992an,Sint:1993un}, 
developed initially to produce a precise determination of the running 
coupling \cite{alpha:su3,DellaMorte:2004bc}, has proved useful also in
phenomenological contexts, like the RG running of the quark mass
\cite{mbar:pap1,Guagnelli:2004za,DellaMorte:2005kg,mbar:pap3}, the
computation of moments of structure functions
\cite{StF}, the evolution of the
static-light axial current \cite{Heitger:2003xg,DellaMorte:2006sv} and
the computation of the Kaon $B$-parameter
\cite{Guagnelli:2005zc,Dimopoulos:2006dm,Dimopoulos:2007cn}. In this
framework, the operator running can be determined by computing the
so-called step scaling function (SSF) for a wide range of renormalized
couplings, which extend from perturbative to non-perturbative
regimes. The SSF itself is determined through a recursive finite-size
scaling procedure, which provides a step-wise construction of the
solution to the Callan-Symanzik equation. Through a sequence of Monte
Carlo simulations at different lattice spacings the latter is obtained
in the continuum limit.

\parbreak The implementation of the non-perturbative renormalization
programme in the framework of the SF is usually split into two parts.
The first is the determination of the scale dependence of the relevant
operators from low to high scales in an SF scheme, which yields
universal, regularization-independent relations between
renormalization group invariant (RGI) operators and their counterparts
in the SF scheme. The second part is the matching between the
operators in the chosen SF scheme and the lattice-regularized theory. 
This is achieved by computing the relevant renormalization factors at a 
fixed low-energy hadronic scale $\mu_{\rm had}$ for several values of the
lattice spacing. The combination of the renormalization factors with
the regularization-independent part yields the total matching between
the bare lattice operators and the RGI ones. In this paper we report
on the determination of the total renormalization factor in quenched
QCD, with the heavy quarks treated in the static approximation
and the light quarks discretized according to the {\rm O}(a) improved 
Wilson action.
\parbreak In order to provide useful input for phenomenology, lattice
determinations of $B$-parameters must have an accuracy at the level of
a few percent. Thus, to avoid being dominated by the numerical
uncertainty in the renormalization factor, we aim for a target
precision of the RGI constants within 1.5-2\% 
in this work. It is well known that Monte Carlo
simulations including static fermions are plagued by a deterioration
of the numerical signal. As shown in \cite{DellaMorte:2005yc}, this
problem can be overcome through the adoption of statistically improved
actions. An analysis of the signal-to-noise ratio shows that achieving
a relative uncertainty around $1\%$ in the continuum limit of the SSF
is unattainable when the naive discretization of the Eichten-Hill (EH)
fermions is employed, especially in the deeply non-perturbative
regime. The use of different lattice discretizations allows to obtain
independent determinations of the SSF at finite lattice
spacing. Universality of the continuum limit then imposes the
constraint that results from different discretizations extrapolate to
a common value at vanishing lattice spacing. This fact can be
exploited in order to constrain fits corresponding to different 
discretizations, so to reduce the systematic uncertainty.
\parbreak The paper is organized as follows. In
sect.~\ref{sec_operators} we introduce the multiplicatively
renormalizable operator basis
$\left\{{\cQ'}_k^\lp\right\}_{k=1,\dots,4}$. The RG equation, its
formal solution and the strategy used for the reconstruction of the
operator scale evolution in various SF schemes are reviewed in
sect.~\ref{sec_running}. Details concerning the lattice formulation
and the Monte Carlo simulations are reported in
sect.~\ref{sec_lattice}. Sect.~\ref{sec_results} is devoted to the
analysis of the numerical results. Here we present a discussion of
the noise-to-signal ratios observed in our simulations, the continuum
extrapolation of the SSF, the RG running in the continuum limit and
the connection to the hadronic observables. Conclusions are drawn in
sect.~\ref{sec_concl}. Tables and plots have been collected in 
appendix A.

\section{Static-light four-fermion operators \label{sec_operators}}

Here we briefly review the definition of the operator basis used in
our calculation. For full details, see sect.~2 of
ref.\,\cite{Palombi:2006pu}.
\parbreak
We consider a theory with a light quark sector consisting of two
massless ${\rm O}(a)$ improved Wilson-type fermions $(\psi_1,\psi_2)$
and a heavy quark, represented by a pair of static fields
$(\psi_h,\psi_{\bar h})$, which propagate respectively forward and
backward in time. We are interested in $\Delta B=2$ static-light
four-fermion operators. These are generically defined via
\begin{equation}
\label{genop}
\cO^\pm_{\Gamma_1\Gamma_2} =
    \dfrac{1}{2}\left[(\bar\psi_{h}\Gamma_1\psi_1)(\bar\psi_{\bar
    h}\Gamma_2\psi_2) \pm (\bar\psi_{h}\Gamma_1\psi_2)(\bar\psi_{\bar
    h}\Gamma_2\psi_1)\right]\ ,
\end{equation}
where $\Gamma_{1,2}$ are Dirac matrices, and the notation
\begin{equation}
\cO^\pm_{\Gamma_1\Gamma_2\ \pm\ \Gamma_3\Gamma_4} \equiv
\cO^\pm_{\Gamma_1\Gamma_2} \pm \cO^\pm_{\Gamma_3\Gamma_4} 
\end{equation}
is adopted. Our attention will be restricted to the subset of the
above operators which are odd under parity and are eigenvectors of the
flavour exchange symmetry
$\left\{\cS:\psi_1\leftrightarrow\psi_2\right\}$ with positive
eigenvalue. The operator basis commonly used in the literature is
\begin{equation}
  \label{stdbasis}
  \left({\cQ}_1^\lp,{\cQ}_2^\lp,{\cQ}_3^\lp,{\cQ}_4^\lp\right) =
  \left(\OVApAV,\OSPpPS,\OVAmAV,\OSPmPS\right)\ . 
\end{equation}
Note that the tensor structure ${\rm T\tilde T}$ is redundant in the
static approximation. The above operator basis exhibits a non-trivial
mixing pattern under renormalization, which makes it unsuitable to a
non-perturbative numerical study of the RG running. As shown in
\cite{Palombi:2006pu}, the mixing can be fully disentangled by taking
appropriate linear combinations of the $\cQ_k^\lp$'s with integer
coefficients, namely
\begin{equation}
  \label{diagbasis}
\left({\cQ'}_1^\lp,{\cQ'}_2^\lp,{\cQ'}_3^\lp,{\cQ'}_4^\lp\right) =
\left({\cQ}_1^\lp,{\cQ}_1^\lp + 4 {\cQ}_2^\lp,{\cQ}_3^\lp + 2
{\cQ}_4^\lp,{\cQ}_3^\lp - 2 {\cQ}_4^\lp\right) \ .
\end{equation}
The operators ${\cQ'}_k^\lp$ renormalize purely multiplicatively. The
existence of a rearrangement of the standard operators, which yields
multiplicative renormalizability without the need for a fine tuning of
the mixing coefficients with the bare coupling, is a consequence of
the heavy quark spin symmetry characterizing the effective static
field theory. It is therefore peculiar to the $\Delta B=2$ parity-odd
four-fermion operators in HQET.

\section{Renormalization group running \label{sec_running}}

In order to prepare the ground for our study of the scale evolution 
of the operators ${\cQ'}_k^\lp$, some basic concepts of 
the RG theory are briefly reviewed. A sketch of the computational strategy
for the numerical reconstruction of the non-perturbative RG running in the SF
scheme is then depicted. 

\subsection{Callan-Symanzik equation}

The scale evolution of the operators provided by Eq.~(\ref{diagbasis})
is governed by a set of scalar Callan-Symanzik equations,
\begin{equation}
\label{CSeq}
\left[\mu\frac{\partial}{\partial\mu} + \beta\frac{\partial}{\partial g_\rR}
+ \tau\sum_{j=1}^{N_{\rm f}}m_{\rR,j}\frac{\partial}{\partial m_{\rR,j}} - 
{\gamma'}^\lp_k\right]\left({\cQ'}_k^\lp\right)_\rR = 0\ ,
\end{equation}
where $k=1,\ldots,4$, and the renormalized operator is related to the
bare lattice one through
\begin{equation}
\left({\cQ'}_k^\lp\right)_\rR(\mu) = \lim_{a\to 0}
     {\cZ'}_k^\lp(g_0,a\mu) {\cQ'}_k^\lp(a) \ . 
\end{equation}
Here $g_0$ denotes the bare gauge coupling. If a mass-independent 
renormalization scheme is adopted, as assumed in the following, the 
RG functions $\beta$, $\tau$ and ${\gamma'}_k^\lp$ depend only upon 
the coupling. In particular, $\beta(g)$ and $\tau(g)$ control the 
running of the renormalized parameters $\bar g(\mu)$ and 
$\overline{m}_j(\mu)$ through the RG equations
\begin{equation}
\mu\frac{\partial\bar{g}}{\partial \mu}  = \beta(\bar{g}) \ , \qquad
\mu\frac{\partial \overline{m}_j}{\partial\mu} = 
\tau(\bar g)\overline{m}_j\ ,
\end{equation}
while the anomalous dimension ${\gamma'}^\lp_k(g)$, which provides the
radiative correction to the classical scaling of ${\cQ'}_k^\lp$, is
related to the renormalization constant ${\cZ'}_k^\lp$ via a
logarithmic derivative,
\begin{align}
{\gamma'}^\lp_k\left(\bar g(\mu)\right) & = \lim_{a\to0}\left(\mu\frac{\partial}{\partial\mu}{\cZ'}_k^\lp(g_0,a\mu)\right) 
{\cZ'}_k^\lp(g_0,a\mu)^{-1}\ .
\end{align}
We emphasize that $\beta$, $\tau$ and ${\gamma'}_k^\lp$ are
non-perturbatively defined functions. Their dependence upon the
coupling constant in the short-distance regime is expected to be
asymptotically described by the first terms of the perturbative
expansions
\begin{align}
\beta(g)     & = -g^3\left[b_0 + b_1g^2 + b_2g^4 + O(g^6)\right] \ , \\[1.8ex]
\tau(g)      & = -g^2\left[d_0 + d_1g^2 + O(g^4)\right] \ , \\[1.2ex]
{\gamma'}^\lp_k(g) & = -g^2\left[{\gamma'}_k^{\lp;(0)} + {\gamma'}_k^{\lp;(1)}g^2 + O(g^4)\right] \ .
\end{align}
The universality of the lowest order coefficients can be demonstrated
by relating the Callan-Symanzik equations corresponding to different
renormalization schemes. In particular, the leading order (LO)
coefficients $b_0$ and $d_0$, and the next-to-leading order (NLO) one
$b_1$ are found~to~be
\begin{align}
b_0 & = (4\pi)^{-2}\left\{\tfrac{11}{3}N - \tfrac{2}{3}N_{\rm
  f}\right\}\ , \\[1.2ex] 
b_1 & = (4\pi)^{-4}\left\{\tfrac{34}{3}N^2 -
  \left(\tfrac{13}{3}N - N^{-1}\right)N_{\rm f}\right\}\ , \\[1.2ex]
d_0 & = (4\pi)^{-2}\left\{3N-3N^{-1}\right\} \ .
\end{align}
in all renormalization schemes. The LO coefficients
${\gamma'}_k^{\lp;(0)}$ of the anomalous dimensions of four-fermion
operators are universal as well. Their values have been obtained in
\cite{Palombi:2006pu} by rotating the LO coefficient of the anomalous
dimension matrix in the operator basis ${\cQ}_k^\lp$, originally
computed in \cite{Flynn:1990qz,Gimenez:1998mw}, to the diagonal basis ${\cQ'}_k^\lp$. These
coefficients read
\begin{align}
{\gamma'}_1^{+;(0)} & = -(4\pi)^{-2}\left(3N - 3N^{-1}\right)\ ,   \\[1.2ex]
{\gamma'}_2^{+;(0)} & = -(4\pi)^{-2}\left(3N-4-7N^{-1}\right)\ ,  \\[1.2ex]
{\gamma'}_3^{+;(0)} & = -(4\pi)^{-2}\left(3N+3-6N^{-1}\right)\ ,  \\[1.2ex]
{\gamma'}_4^{+;(0)} & = -(4\pi)^{-2}\left(3N-3-6N^{-1}\right)\ .
\end{align}
\parbreak
The formal solution of the Callan-Symanzik equation relates the
scheme-dependent RG running operator
$\left({\cQ'}_k^\lp\right)_\rR(\mu)$ to the renormalization group
invariant one $\left({\cQ'}_k^\lp\right)_\RGI$,
\begin{equation}
\label{RGsol}
\left({\cQ'}_k^\lp\right)_\RGI  = \left({\cQ'}_k^\lp\right)_\rR(\mu)
\left[\frac{\bar g^2(\mu)}{4\pi}\right]^{-{{\gamma'}_k^{\lp;(0)}}/{2b_0}}
\exp\left\{-\int_0^{\bar g(\mu)}dg\left(\frac{{\gamma'}^\lp_k(g)}{\beta(g)} - \frac{{\gamma'}_k^{\lp;(0)}}{b_0g}\right)\right\}\ .
\end{equation}
From a mathematical point of view, the RGI operator can be interpreted
as the ``integration constant'' of the solution of the Callan-Symanzik
equation. As such, it is uniquely defined up to an overall
scale-independent factor. In Eq.~(\ref{RGsol}) we have adopted the
normalization usually employed with four-fermion operators. The RGI
operator can be easily shown to be independent of the renormalization
scheme. Note that all the scale dependence is carried by a factor,
\begin{equation}
\label{evfac}
{\hat c}'^{\lp}_{\:k}(\mu) = \left[\frac{\bar
    g^2(\mu)}{4\pi}\right]^{-{\gamma'}_k^{\lp;(0)}/2b_0} 
\exp\left\{-\int_0^{\bar
  g(\mu)}dg\left(\frac{{\gamma'}^\lp_k(g)}{\beta(g)} -
\frac{{\gamma'}_k^{\lp;(0)}}{b_0g}\right)\right\}\ , 
\end{equation}
which represents the integration of the RG functions $\beta(g)$ and
${\gamma'}^\lp_k(g)$ in the whole range of renormalization scales from
$\mu$ to infinity. This integral receives perturbative contributions
in the region where $\bar g^2(\mu) \ll 1$. The total amount of
non-perturbative contributions depends on how deeply in the
non-perturbative regime the renormalization scale $\mu$ is placed and
on the rate of convergence of perturbation theory at the scale $\mu$
in the chosen renormalization scheme.

\subsection{Step scaling functions and total renormalization factor}

The computation of the evolution factor ${\hat{c}}'^{\lp}_{\:k}(\mu)$
requires full knowledge of the RG functions over a large range of
scales. Numerical simulations can provide an insight into the
non-perturbative region, but for that purpose Eq.~(\ref{evfac}) is of
little practical use. We shall now describe how the scale evolution
can be determined non-perturbatively from low energies, corresponding
to typical hadronic scales, to high energies, where the coupling is
sufficiently small to make contact with perturbation theory.
\parbreak
Our task is to compute the proportionality factor between renormalized
operators at a low-energy hadronic scale $\mu_{\rm{had}}$ and their
counterparts at the scale $\mu$, i.e.
\begin{equation}
\label{RGconn}
  \left({\cQ'}_k^\lp\right)_\rR(\mu) =
   {U'}^\lp_k(\mu,\mu_{\rm had})\left({\cQ'}_k^\lp
  \right)_\rR(\mu_{\rm{had}}) \ . 
\end{equation}
The renormalization is multiplicative, and hence ${U'}^\lp_k$ is given
by the ratio
\begin{equation}
\label{factorization}
  {U'}^\lp_k(\mu,\mu_{\rm had})  =
  {\hat{c}}'^{\lp}_{\:k}(\mu_{\rm had})/{\hat{c}}'^{\lp}_{\:k}(\mu)\ .
\end{equation}
Typically, we will think of the scale $\mu$ to lie in the ultraviolet,
such that $\mu\gg\mu_{\rm had}$. Since it is difficult to accommodate
scales that differ by orders of magnitude in a single lattice
calculation, it is useful to factorize the evolution and adopt a
recursive approach. The so-called step scaling functions (SSFs)
$\sigma_k^{+}$ and $\sigma$ describe the change in the operators and
the gauge coupling, respectively, when the energy scale $\mu$ is
decreased by a factor~2, i.e.
\begin{align}
  & \sigma(u) = \bar{g}^2(\mu/2)\ ,\qquad u\equiv\bar{g}^2(\mu)\ ; \nonumber \\[2.0ex]
  & \sigma_k^{+}(u) = \frac{\hat{c}'^{\lp}_{\:k}(\mu/2)}{\hat{c}'^{\lp}_{\:k}(\mu)}\ .
\end{align}
In sect.~\ref{SFrunning} we shall sketch how $\sigma_k^{+}$ and
$\sigma$ can be computed for a sequence of
couplings~$u_i,\,i=0,1,2,\ldots$ in lattice simulations. For the moment we
simply state that the relation between operators renormalized at
scales $\mu_{\rm had}$ and $2^n\mu_{\rm had}$ is obtained from the
product of SSFs via
\begin{equation}
\label{eq:NPevol}
  {U'}^\lp_k(2^n\mu_{\rm had},\mu_{\rm had}) = \left\{\prod_{i=0}^{n-1}
  {\sigma}^\lp_k(u_i)\right\}^{-1}\ ,
  \quad u_i=\bar{g}^2(2^{(i+1)}\mu_{\rm had})\ .
\end{equation}
If $\mu_{\rm had}$ is taken to be a few hundreds of MeV, it is safe to
assume that $2^n\mu_{\rm had}$ lies in a regime where perturbation
theory can be applied, provided that one succeeds in computing the
SSFs for a sufficiently large number of steps. In our numerical
determination described in sect.~\ref{sec_results} we have used $n=8$, 
and thus we could trace the evolution non-perturbatively over three 
orders of magnitude.
\parbreak
Assuming that $\mu_{\rm pt}\equiv 2^n\mu_{\rm had}$ is large enough,
one can evaluate $\hat{c}'^{\lp}_{\:k}(\mu_{\rm pt})$ by inserting the
perturbative expressions for the anomalous dimensions and the
$\beta$-function into Eq.~(\ref{evfac}). The relation between the RGI
operators and their counterparts at the hadronic scale is thus given
by
\begin{equation}
  \left({\cQ'}_k^\lp\right)_\RGI = {\hat{c}}'^{\lp}_{\:k}(\mu_{\rm pt})
    {U'}^\lp_k(\mu_{\rm pt},\mu_{\rm had})
    \left({\cQ'}_k^\lp\right)_\rR(\mu_{\rm had})\ .
\end{equation}
It remains to specify the total renormalization factor
$\hat{Z}'^\lp_{\ k,\RGI}$ which links the RGI operator to the bare
operator ${\cQ'}_k^\lp(a)$ on the lattice via
\begin{equation}
  \left({\cQ'}_k^\lp\right)_\RGI = \hat{Z}'^\lp_{\ k,\RGI}(g_0)
  {\cQ'}_k^\lp(a)\ ,
\end{equation}
where the total renormalization factor is given by the product
\begin{equation}
\label{eq:totalren}
  \hat{Z}'^\lp_{\ k,\RGI}(g_0)={\hat{c}}'^{\lp}_{\:k}(\mu_{\rm pt})
    {U'}^\lp_k(\mu_{\rm pt},\mu_{\rm had})
    {\cZ}'^\lp_{\:k}(g_0,a\mu_{\rm had})\ .
\end{equation}
The factor ${\cZ'}_k^\lp(g_0,a\mu_{\rm had})$ must be determined for
each operator in a lattice simulation at fixed $\mu_{\rm had}$ for a
range of bare couplings, using suitable renormalization conditions. We
stress that the combination ${\hat{c}}'^{\lp}_{\:k}(\mu_{\rm pt})
{U'}^\lp_k(\mu_{\rm pt},\mu_{\rm had})$ represents the universal,
regularization-independent contribution to
$\hat{Z}'^\lp_{\ k,\RGI}$. Finally, we note that all reference to the
scales $\mu_{\rm pt}$ and $\mu_{\rm had}$ drops out in the total
renormalization factor.

\subsection{Scale evolution in the SF\label{SFrunning}}

The non-perturbative renormalization of local composite operators via
the Schr\"odinger functional has become a standard method. The SF
scheme is based on the formulation of QCD in a finite space-time
volume $T\times L^3$, with periodic spatial boundary conditions and
Dirichlet boundary conditions at Euclidean times $x_0=0,T$.
\cite{Luscher:1992an,Sint:1993un}. By imposing suitable
renormalization conditions at vanishing quark mass and by choosing a
particular aspect ratio $T/L$, the box size $L$ remains the only scale
in the formulation. The dependence of composite operators and the
gauge coupling on the renormalization scale can thus be probed by
changing the volume. In particular, the step scaling functions for a
variety of operators can be computed via recursive finite-size
scaling, ranging over several orders of magnitude in the physical box
size.
\parbreak
In order to fully specify our adopted finite-volume scheme, we have
set the aspect ratio to $T/L=1$. Furthermore, as in
ref.~\cite{mbar:pap1} we have imposed periodic spatial boundary
conditions up to a phase $\theta=0.5$ and evaluated the
renormalization conditions for vanishing background field.
\parbreak
Renormalization conditions for our four-quark operators in the SF
scheme are defined in sect.~3 of ref.~\cite{Palombi:2006pu}, to which
the reader is referred for full details. In particular, the relevant
renormalization factors ${\cZ'}^{\lp}_{k}$ are given in terms of
suitable correlation functions of the operators ${\cQ'}_k^\lp$ (see
Eq.~(3.16) of \cite{Palombi:2006pu}). Note that in
\cite{Palombi:2006pu} our notation for the renormalization constants
and the corresponding SSFs is supplemented by two additional indices,
e.g. ${\cZ'}^{\lp;(s)}_{k;\alpha}$. The index $s=1,\ldots,5$
enumerates the different boundary Dirac structures\footnote{See
Eqs.~(3.4)--(3.8) of \cite{Palombi:2006pu}.}  which can be used in
order to probe the four-fermion operators ${\cQ'}_k^\lp$; the index
$\alpha=0,1/2$ distinguishes different combinations of pseudo-scalar
and vector boundary-to-boundary bilinear correlators, used to
remove the additional divergencies introduced by the boundary
sources\footnote{See Eqs.~(3.11)--(3.15)
of~\cite{Palombi:2006pu}.}. However, for notational clarity we shall
drop the additional indices in the following.
\parbreak
For each combination of~$s$ and $\alpha$ we compute the lattice SSFs
of the operators ${\cQ'}^{\lp}_k$, defined as
\begin{equation}
  \Sigma_{k}^{\lp}(u,a/L) = \left.
  \frac{{\cZ'}^{\lp}_{k}(g_0,a/2L)}{{\cZ'}^{\lp}_{k}(g_0,a/L)}
  \right|_{m=0,\,\gbar^2_{\rm SF}(L)=u} \ ,
\end{equation}
i.e. the SSFs are evaluated in the chiral limit, $m(g_0)=0$, (where
$m$ is the PCAC quark mass defined following ref.~\cite{mbar:pap1}),
for a given lattice size $L/a$ and at fixed renormalized SF coupling
$\gbar^2_{\rm\scriptscriptstyle SF}(L)=u,\,\mu=1/L$. The lattice SSFs 
$\Sigma_{k}^{\lp}$ depend not only on the definition of the 
renormalization scheme, but also on the details of the lattice 
regularization. They have, however, a well defined continuum limit, 
viz.
\begin{equation}
  \sigma_{k}^{\lp}(u) = \lim_{a \to 0}\Sigma_{k}^{\lp}(u,a/L) \ .
\end{equation}
Thus, at each fixed value of the renormalized coupling, the SSFs in
the continuum limit are obtained by computing
$\Sigma_{k}^{\lp}(u,a/L)$ for several values of the lattice spacing
and performing an extrapolation to vanishing lattice spacing.
\parbreak
Our task is the determination of the scale evolution factor
${U'}^\lp_k$ of Eq.~(\ref{eq:NPevol}) for $\mu_{\rm had}=1/(2L_{\rm
max})$, where the scale $L_{\rm max}$ is implicitly defined through
\begin{equation}
   \bar g^2_{\rm\scriptscriptstyle SF}(L_{\rm max})=3.48\ .
\end{equation}
This value of the coupling corresponds to $L_{\rm max}/r_0=0.738(16)$
\cite{Necco:2001xg}, which for $r_0=0.5$\,fm translates into
$\mu_{\rm had}\approx270$\,MeV. The sequence of couplings
\begin{equation}
   u_i=\bar g^2_{\rm\scriptscriptstyle SF}(2^{-i}L_{\rm max})\ ,\qquad
   i=0,1,2,\ldots 
\end{equation}
is computed by solving the recursion
\begin{equation}
\label{eq:coupevol}
  u_0 = 3.480\ , \quad \sigma(u_{l+1})=u_l\ .
\end{equation}
The SSF of the coupling $\sigma(u)$ has been calculated in the
quenched approximation in~\cite{alpha:su3,mbar:pap1}. The SSFs of the
four-fermion operators can then be evaluated for the sequence of
couplings, $u_l,\,l=0,1,2,\ldots$, and Eq.~(\ref{eq:NPevol}) yields
the RG evolution between the hadronic scale $\mu_{\rm had}=1/(2L_{\rm
max})$ and the high-energy scale $\mu=2^{n-1}/L_{\rm max}$.
\parbreak
As described below in sect.~\ref{sec:RGrunning}, in practice we fit
the data for each SSF to a polynomial and use the resulting fit
functions in the recursions of Eqs.~(\ref{eq:coupevol})
and~(\ref{eq:NPevol}).

\section{Lattice setup \label{sec_lattice}}

\subsection{Discretization of light and heavy quarks}

As previously stated, light quarks are discretized in this work
according to the Wilson prescription with $\Oa$ Symanzik
improvement. The general concept how to implement $\Oa$ improvement in
the SF has been presented in
refs.~\cite{Luscher:1992an,Luscher:1996sc}. As usual, the improvement
of the Wilson action is achieved by adding the standard
Sheikholeslami-Wohlert term \cite{Sheikholeslami:1985ij}.
\parbreak
Field theories defined in finite volume with boundaries, such as the
SF of QCD, require that suitable boundary counterterms be included as
well, in order to fully cancel ${\rm O}(a)$ lattice artefacts. The
particular realization of the SF of
refs.\,\cite{Luscher:1992an,Luscher:1996sc}, which we adopt in this
paper, lists two relevant counterterms, multiplied by the improvement
coefficients $c_{\rm t}(g_0^2)-1$ and $\tilde c_{\rm t}(g_0^2)-1$,
respectively.
\parbreak
The improvement coefficient $c_{\rm sw}$ has been computed
non-per\-tur\-ba\-tively in the quenched approximation for a range of
values of the bare coupling $g_0$ \cite{Luscher:1996ug} and is
parameterized by the interpolating formula
\begin{equation}
\label{eq:csw}
c_{\rm sw}(g_0^2) = \frac{1-0.656g_0^2 - 0.152g_0^4 -
  0.054g_0^6}{1-0.922g_0^2}\ .
\end{equation}
By contrast, the coefficients $c_{\rm t}$ and $\tilde c_{\rm t}$ are known only in
perturbation theory to NLO~\cite{Bode:1998hd} and
LO~\cite{Luscher:1996vw} respectively:
\begin{align}
\label{eq:ctPT}
c_{\rm t}(g_0^2) & = 1 - 0.089 g_0^2 - 0.030 g_0^4 \ , \\[1.65ex]
\label{eq:cttPT}
\tilde c_{\rm t}(g_0^2) & = 1 - 0.018 g_0^2 \ .
\end{align}
\parbreak
Heavy quarks are treated in the static approximation. The original
lattice action, first derived by Eichten and Hill in
\cite{Eichten:1989zv}, has been subsequently generalized in
\cite{DellaMorte:2005yc}, in order to improve the signal-to-noise
ratio of static-light correlators at large time separations. Following
this approach, we write it in the form
\begin{equation}
\label{staticaction}
S^{\rm stat}[\psi_{h},\bar \psi_{h},\psi_{\bar h},\bar\psi_{\bar h},U] 
= a^4\sum_x\left[\bar\psi_h(x)D_0^{{\rm W}*}\psi_h(x)-\bar\psi_{\bar h}
(x)D_0^{{\rm W}}\psi_{\bar h}(x)\right]\ ,
\end{equation}
where the covariant derivatives are defined according to
\begin{align}
D_0^{\rm W}\psi(x) & = \frac{1}{a}\left[W_0(x)
\psi(x+a\hat 0) - \psi(x)\right]\nonumber\ ,\\[2.0ex]
D_0^{\rm W*}\psi(x) & = \frac{1}{a}\left[\psi(x) - 
W_0^\dagger(x-a\hat 0)\psi(x-a\hat 0)\right] \ .
\end{align}
In order to reproduce the original Eichten-Hill formulation, the
generalized parallel transporter $W_0(x)$ must be replaced by the
temporal gauge link $U_0(x)$. Moreover, the choice of $W_0(x)$ is
constrained by the requirement of keeping the theory in the same
universality class, to guarantee a unique continuum limit. In the
following, we consider four different choices of $W_0(x)$, all
compliant with this requirement, i.e.
\begin{align}
W_0^{\rm EH}(x)   & = U_0(x)\ , \\[2.0ex]
W_0^{\rm APE}(x)  & = V_0(x)\ , \\[2.0ex]
W_0^{\rm HYP1}(x) & = V^{\rm HYP}_0(\vec\alpha, x)\bigr|_{{\vec \alpha} = (0.75,0.6,0.3)} \ , \\[2.0ex]
W_0^{\rm HYP2}(x) & = V^{\rm HYP}_0(\vec\alpha, x)\bigr|_{{\vec \alpha} = (1.0,1.0,0.5)}\ .
\end{align}
In the above definitions $V_0(x)$ denotes the average of the six
staples surrounding the gauge link $U_0(x)$ and $V^{\rm HYP}_0(x)$
represents the temporal HYP link of \cite{Hasenfratz:2001hp}
with the approximate SU(3) projection of
\cite{DellaMorte:2005yc}. Two sets of HYP-smearing coefficients
$\vec\alpha$ are considered, leading to two independent realizations
of the HYP smeared parallel transporter. The static actions so
assembled are automatically $\Oa$ improved, without the need of
time-boundary counterterms, and differ among each other at finite
cutoff by ${\rm O}(a^2)$ terms.
\parbreak 
The $\Oa$ improvement of correlation functions of composite
operators is completed through the inclusion of the appropriate higher
dimension counterterms in the lattice definition of the local
operators. We do not employ operator improvement here, and therefore
we expect that the dominant discretization effects are of $\Oa$. We
note, however, that the correlation functions defined in Eq.~(3.9) of
\cite{Palombi:2006pu} are $\Oa$ tree-level improved, implying that all
$\Oa$ counterterms to the local four-fermion operators vanish at this
order. Thus we are left with discretization errors of order $g_0^2 a$.

\subsection{Simulation details}

In our quenched simulations the bare coupling $g_0$ (equivalently,
$\beta=6/g_0^2$) must be tuned for a given lattice size, in order to
produce a fixed value of the renormalized coupling
$\gbar_{\rm\scriptscriptstyle SF}^2$. Furthermore, renormalization 
conditions for four-quark operators are imposed at vanishing quark 
mass, expressed in terms of the critical hopping parameter, $\hopc$.
\parbreak
The complete set of simulation parameters is reported in the first
four columns of Tables~\ref{tab:tab3} and \ref{tab:tab4}. For each of the
14 values of the renormalized SF coupling mentioned, we have
considered four different lattice resolutions, corresponding to $L/a =
6,8,12,16$. Values of $\beta$ have been tuned at the various lattice
spacings so to have $\bar g^2_{\rm\scriptscriptstyle SF}(L) = u_i$. At
fixed bare coupling we define $\hopc$ as the value where the PCAC
quark mass $m(g_0)$ of ref.~\cite{mbar:pap1}
vanishes. Following~\cite{mbar:pap1}, the computation of $\hopc$ is
done at $\theta = 0$. The amount of statistical samples generated in
the course of the Monte Carlo simulations has been fixed according to
the value of the SF coupling and the lattice spacing, ranging from
${\rm O}(1000-1600)$ independent measurements at the smaller cutoffs
down to ${\rm O}(200-300)$ at the larger ones. In order to keep the
statistical uncertainty of the renormalization constants nearly
constant, an increasing number of samples had to be accumulated at the
larger couplings. Gauge configurations have been produced by
alternating heatbath and overrelaxation steps (in the ratio of $L/2a$
heatbath moves per overrelaxation). On each independent configuration
the Dirac operator has been inverted via the BiCGStab solver with
SSOR-preconditioning \cite{Fischer:1996th,Guagnelli:1999nt}.

\section{Numerical results \label{sec_results}}

A compilation of renormalization factors ${\cZ'}_{k}^{\lp}$ for all
renormalized couplings, lattice spacings, schemes and discretizations
of the static action would easily exceed the size of an ordinary
paper. For the sake of reproducibility, we report in
Tables~\ref{tab:tab3}--\ref{tab:tab10} those corresponding to the
particular case of the HYP2 action and our preferred choice of the 
renormalization schemes, i.e.  $(s,\alpha)=(1,0)$ for 
${\cQ'}^\lp_{1,3,4}$ and $(s,\alpha)=(3,0)$ for ${\cQ'}^\lp_{2}$. 
A complete set of tables and plots is available for download from 
the website \cite{website}.

\subsection{Analysis of the noise-to-signal ratios}

\begin{figure}[!h]
\centering
\epsfig{file=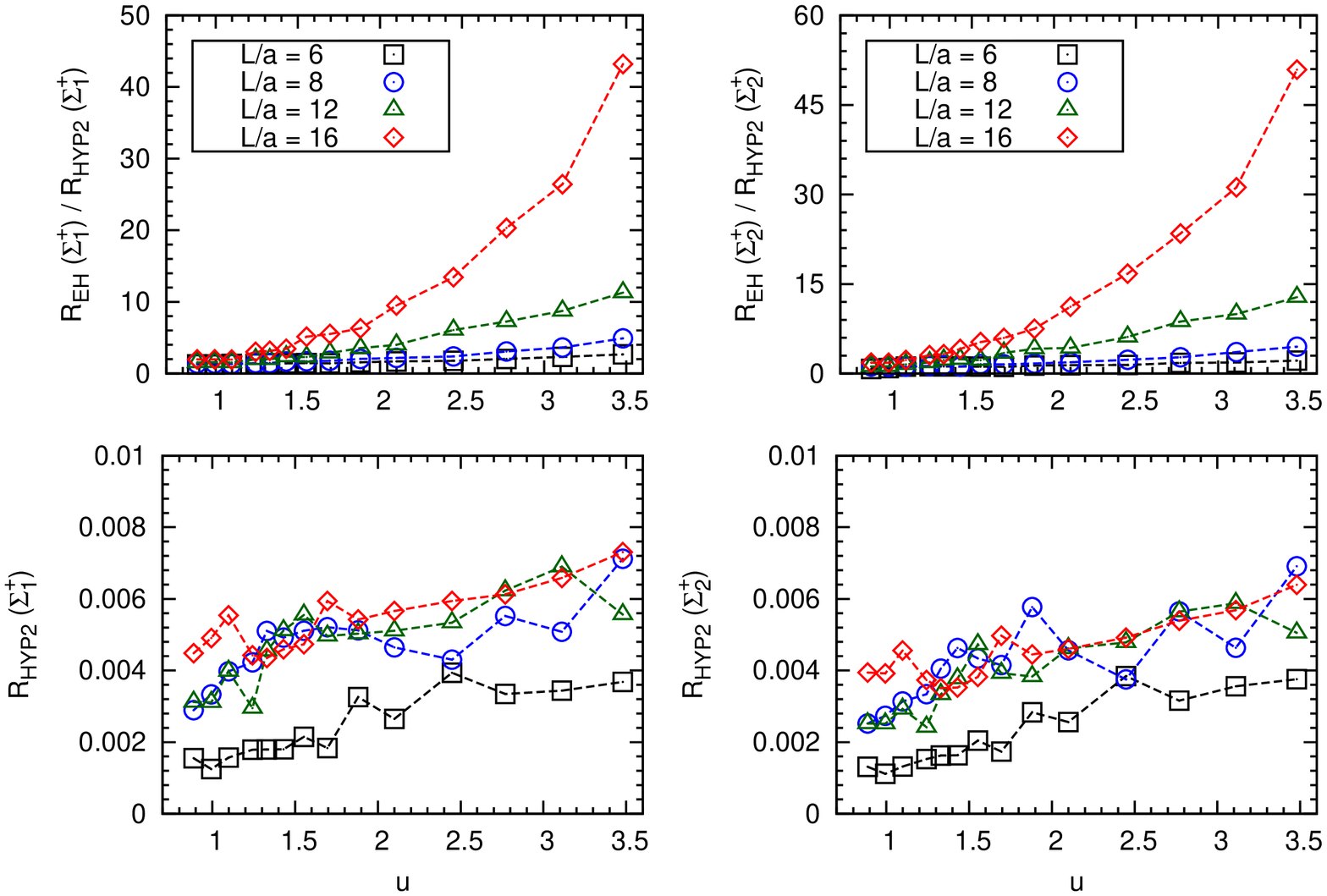,angle=-90,width=13.0cm}
\caption{Comparison of the noise-to-signal ratio of the SSFs
 $\Sigma_{\,1;0}^{\lp;(1)}$ and $\Sigma_{\,2;0}^{\lp;(3)}$
 computed using the EH and HYP2 lattice discretizations of the static
 action.\label{fig:fig1}}
\vskip 0.8cm
\epsfig{file=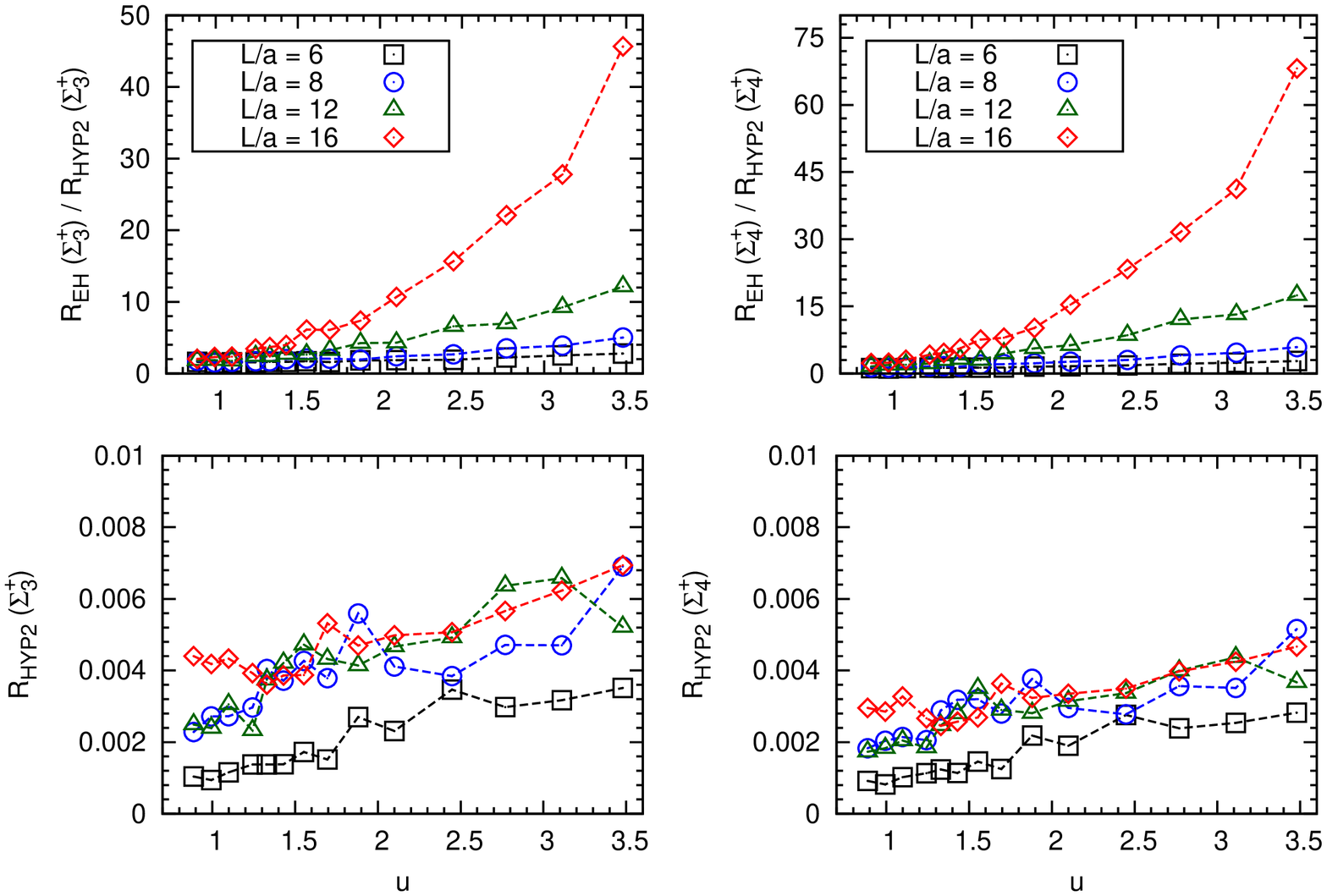,angle=-90,width=13.0cm}
\caption{Same as Fig.\,\protect\ref{fig:fig1}, but for the SSFs
 $\Sigma_{\,3;0}^{\lp;(1)}$ and $\Sigma_{\,4;0}^{\lp;(1)}$.
\label{fig:fig2}
}
\vskip -2.4cm
\end{figure}

A precise determination of the RGI renormalization constants can only
be achieved if the statistical error of the SSF at each simulated
coupling and lattice spacing is kept under control. It is therefore
important to monitor the noise-to-signal ratio
\begin{equation}
R_{\scriptscriptstyle \rm X}(\Sigma_k^\lp) = \frac{\Delta \Sigma_k^\lp}{\Sigma_k^{\lp}}\ , \qquad
{\rm X} = ({\rm EH,APE,HYP1,HYP2})\ ,
\end{equation}
characteristic of the four chosen lattice discretizations of the
static action. Here, $\Delta\Sigma_k^\lp$ denotes the statistical
uncertainty of the SSF $\Sigma_k^\lp$, computed via the jackknife method.
According to\,\cite{DellaMorte:2005yc}, $R_{\scriptscriptstyle\rm X}$ is related to the
value of the binding energy $E_{\rm stat} \sim
\frac{1}{a}e^{(1)}g_0^2~+~\dots$ of the static-light meson, which
diverges linearly in the continuum limit. The leading coefficient
$e^{(1)}$ depends upon the lattice discretization of the static action 
and its value sets the rate of growth of the noise-to-signal ratio: 
a linear reduction of $e^{(1)}$ corresponds indeed to an exponential 
damping of the statistical fluctuations. From our results we deduce 
the general trend
\begin{equation}
R_{\scriptscriptstyle\rm EH} \gg 
R_{\scriptscriptstyle\rm APE} \gtrsim 
R_{\scriptscriptstyle\rm HYP1} \gtrsim 
R_{\scriptscriptstyle\rm HYP2}\ .
\end{equation}
\parbreak
As an example, we compare in Figures~\ref{fig:fig1} and~\ref{fig:fig2}
the noise-to-signal ratio of the SSF of the operators
$\{{\cQ'}_k^\lp\}, k=1,\dots,4$ with EH and HYP2 discretizations in
our preferred choice of the renormalization schemes. Lower
plots in each figure show that $R_{\scriptscriptstyle\rm HYP2}$ is almost constant
against variations in the renormalized coupling and always lower than~$1\%$. 
Moreover, it never increases by more than a factor 4 going from the 
coarsest to the finest lattice resolution at fixed coupling. This picture 
is completely reversed when looking at the EH discretization, as shown 
in the upper plots. Here, a clear increase of the noise-to-signal ratio
with the SF coupling and also with the lattice spacing is observed. In
practice, the SSF has an acceptable uncertainty only in the
perturbative region, i.e. for $u\lesssim\;1$.
\parbreak
A comparison of the noise-to-signal ratio for different operators
shows that in our preferred schemes $\Sigma^\lp_1$, $\Sigma^\lp_2$ and 
$\Sigma^\lp_3$ are slightly noisier than $\Sigma^\lp_4$ with the HYP2 
action, in contrast to the EH one. Since the simulations with the EH 
action are practically unusable for our ultimate aims, the prevailing 
pattern is the one observed with the HYP2 discretization and will be 
reflected in the final statistical error of the RGI renormalization 
constants of the various operators.

\subsection{Continuum extrapolation of the step scaling functions}

The lattice SSFs $\Sigma_{k}^{\lp}$ must be extrapolated to the
continuum limit (i.e. to vanishing $a/L$) at fixed renormalized 
gauge coupling in order to obtain their continuum counterparts 
$\sigma_{k}^{\lp}$. Since the four-fermion operators have not 
been improved, we expect the dominant discretization effects to 
be $\Oa$; thus, our data should exhibit a linear behaviour in 
$a/L$. For every combination of $(s,\alpha)$ we have therefore 
fitted to the {\it ansatz}
\begin{gather}
\Sigma_{k}^{\lp}(u,a/L) =
\sigma_{k}^{\lp}(u) + \rho(u)\,(a/L) \ .
\end{gather}
Fits have been performed using either the whole available set of
values of $L/a$ or, alternatively, without taking into account the
data at $L/a=6$, which may be subject to higher-order lattice artefacts.
\parbreak
Following the spirit of refs.~\cite{StF,Guagnelli:2004za,Guagnelli:2005zc}, one could perform a combined
fit of the data corresponding to the actions APE,
HYP1 and HYP2, all of which have comparable noise-to-signal ratios in
the range of lattice parameters covered in this work. However, 
data obtained with the above actions differ noticeably only -- if ever -- at
$L/a=6$, and are very strongly correlated. As a consequence, a
combined continuum extrapolation affects only marginally the result
coming from the best choice of the action, i.e. HYP2, with a reduction of 
the relative error of $\sigma_{k}^{\lp}(u)$ at the level of a few 
percent. Thus, without any loss, we will consider only the HYP2 data 
from now on.
\parbreak
Fit results can be summarized as follows:

\begin{enumerate}

\item[(i)] the typical statistical accuracy of our results for
$\sigma_{k}^{\lp}$ ranges from $\sim 0.5\%$ relative error, for the
weakest couplings and fits that keep $L/a=6$, to $\sim 1.5\%$ relative
error at the maximum value of $u$, for fits that discard $L/a=6$.
When $L/a=6$ is dropped, the fitted values of $\rho$ are always essentially
compatible with zero within the statistical uncertainty for
${\cQ'}^\lp_1$ and ${\cQ'}^\lp_3$, signalling a weak cutoff dependence of
$\Sigma_{k}^{\lp}$. For ${\cQ'}^\lp_2$ and ${\cQ'}^\lp_4$, on the other hand,
they are zero only within two standard deviations at 
$u \gtrsim 2$. In fits containing 
$L/a=6$, non-zero values of $\rho$ are usually obtained for the 
couplings $u \gtrsim 2$ and all the operators;

\item[(ii)] results from three-point and four-point fits are
always compatible within one standard deviation for all operators and
schemes, save for a few exceptions in which the agreement is at the
level of $1.5\sigma$ only;

\item[(iii)] the goodness of fit, expressed by the value of $\chi^2$
per degree of freedom, is mostly around or below~$1$, although it
reaches large values in some cases. This does not depend
systematically on the number of the fitted points or the value of the
coupling. Anyway, given the small number of fitted data points,
$\chi^2/\mbox{d.o.f.}$ for each single fit at fixed value of the 
coupling is a goodness-of-fit criterion of limited value; instead, 
the total $\chi^2/\mbox{d.o.f.}$ (summed over all values of the 
coupling at fixed operator and scheme) is always around or below~$1$, 
reaching maxima of the order of $1.5$.

\end{enumerate}

\begin{figure}[!h]
\centering
\epsfig{file=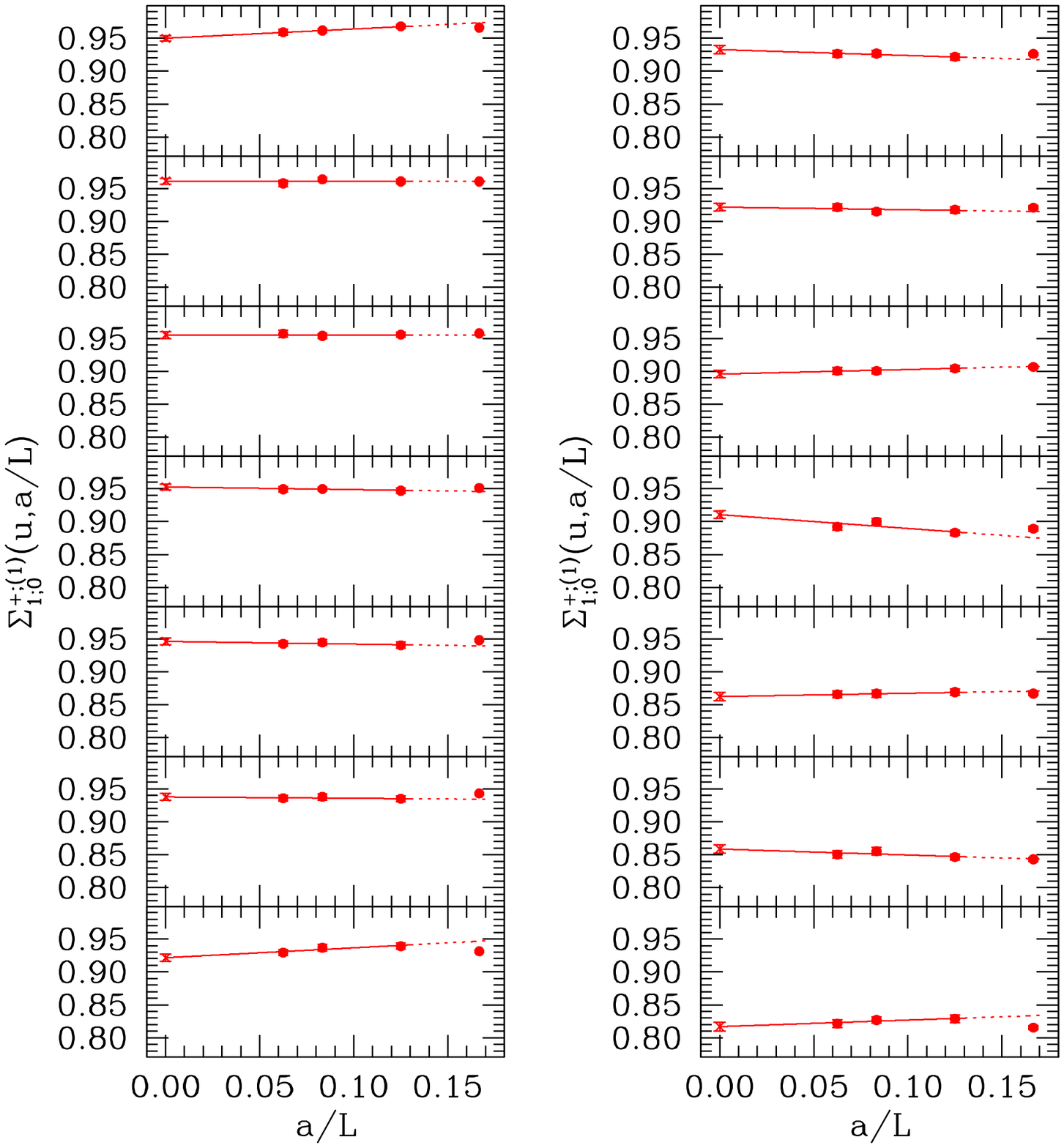,width=14.0cm}
\caption{Continuum limit extrapolation of the SSF
 $\Sigma_{1;0}^{\lp;(1)}$  at various SF renormalized
 couplings with HYP2 lattice discretization of the 
 static action. The SF coupling $u$ increases from top-left to bottom-right,
 according to the first column of Table~\protect\ref{tab:CLe1}.}
 \label{fig:extrap1}
\vskip 0.4cm
\end{figure}

\begin{figure}[!h]
\centering
\epsfig{file=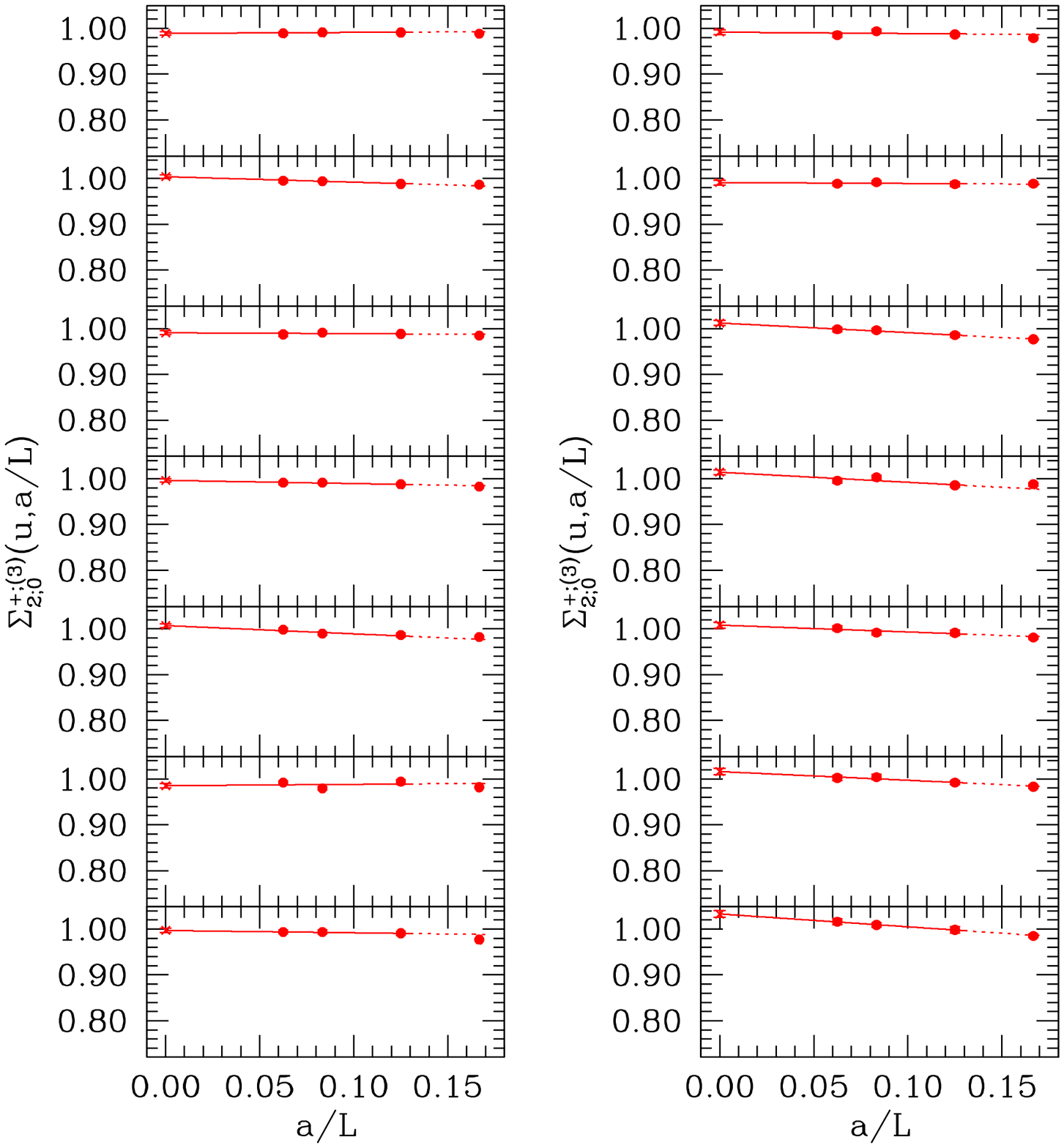,width=14.0cm}
\caption{Continuum limit extrapolation of the SSF
 $\Sigma_{2;0}^{\lp;(3)}$ at various SF renormalized couplings
 with HYP2 lattice discretization of the static
 action. The SF coupling $u$ increases from top-left to bottom-right,
 according to the first column of Table~\protect\ref{tab:CLe1}.}
 \label{fig:extrap2}
\vskip 0.4cm
\end{figure}

\begin{figure}[!h]
\centering
\epsfig{file=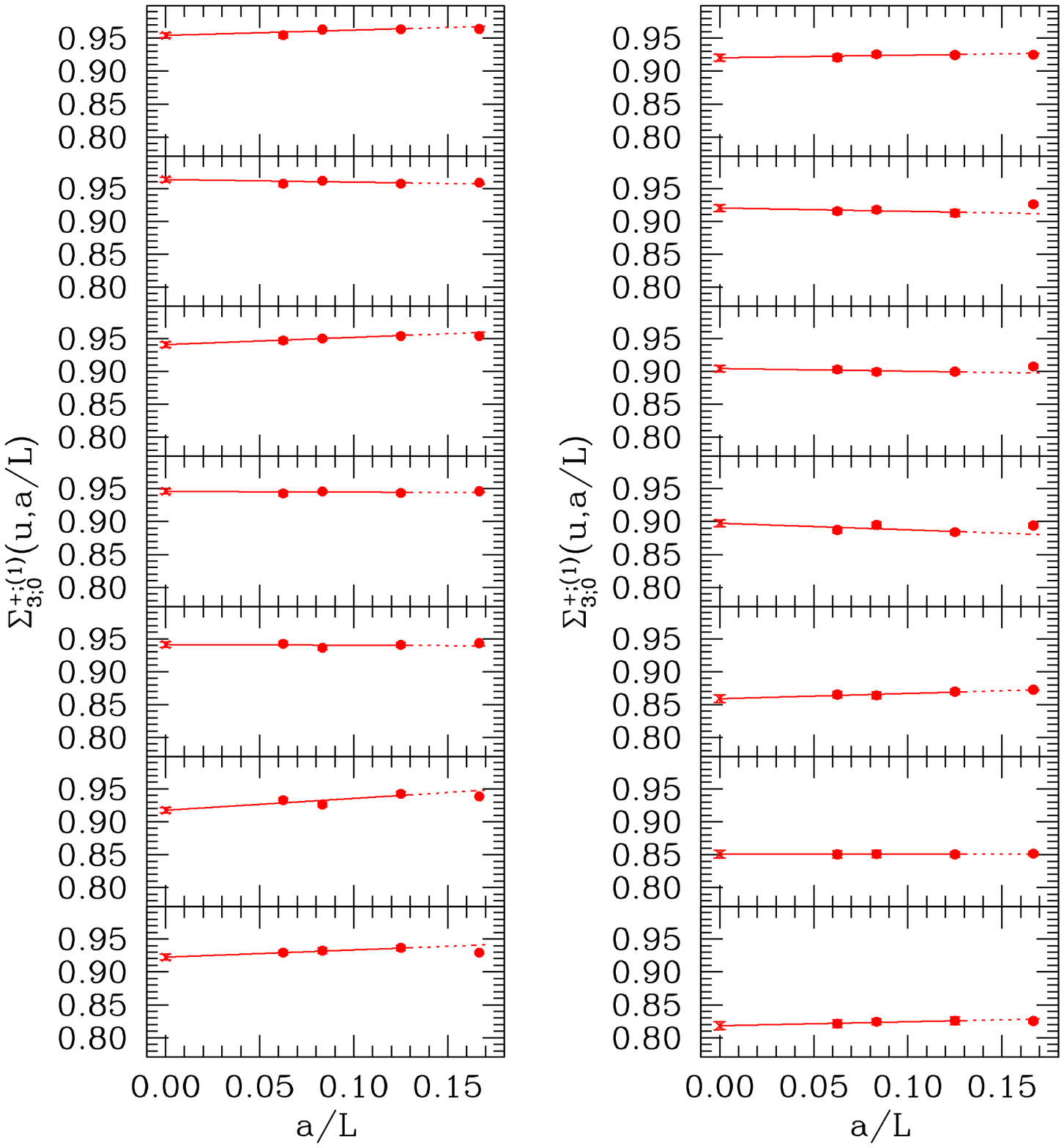,width=14.0cm}
\caption{Continuum limit extrapolation of the SSF
 $\Sigma_{3;0}^{\lp;(1)}$ at various SF renormalized couplings
 with HYP2 lattice discretization of the static
 action. The SF coupling $u$ increases from top-left to bottom-right,
 according to the first column of Table~\protect\ref{tab:CLe1}.}
 \label{fig:extrap3}
\vskip 0.4cm
\end{figure}

\begin{figure}[!h]
\centering
\epsfig{file=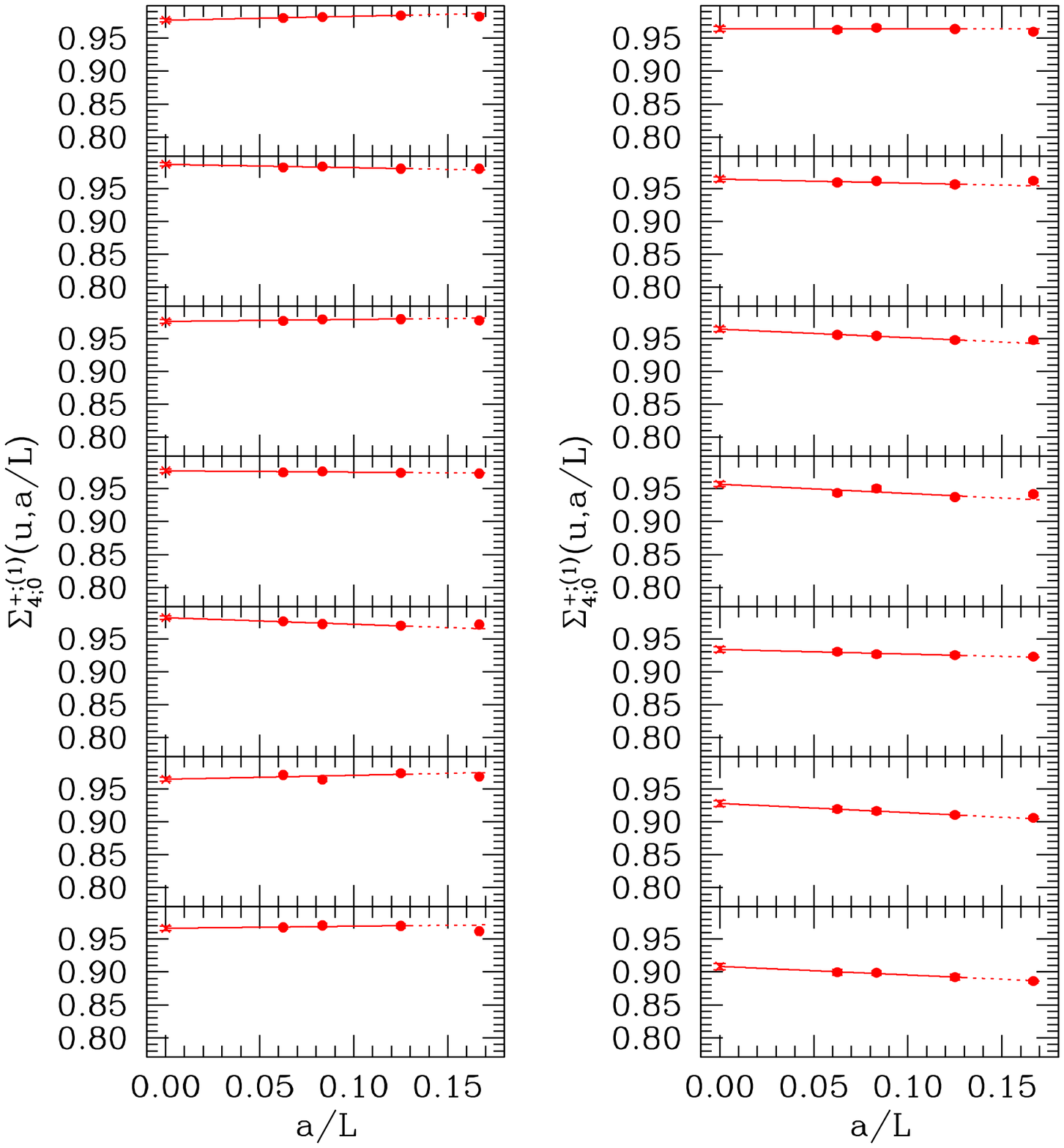,width=14.0cm}
\caption{Continuum limit extrapolation of the SSF
 $\Sigma_{4;0}^{\lp;(1)}$ at various SF renormalized couplings
 with HYP2 lattice discretization of the static
 action. The SF coupling $u$ increases from top-left to bottom-right,
 according to the first column of Table~\protect\ref{tab:CLe1}.}
 \label{fig:extrap4}
\vskip 0.4cm
\end{figure}

Based on this analysis, we conservatively choose linear extrapolations
that do not consider the $L/a=6$ datum to extract our final values of
$\sigma_{k}^{\lp}$. The resulting continuum limit extrapolations are
illustrated in Figs.~\ref{fig:extrap1}--\ref{fig:extrap4} for our 
reference schemes (chosen below). Complete tables with the results are 
available at~\cite{website}.

\subsection{RG running in the continuum limit \label{sec:RGrunning}}

\begin{figure}[!h]
\vspace{285mm}
\includegraphics{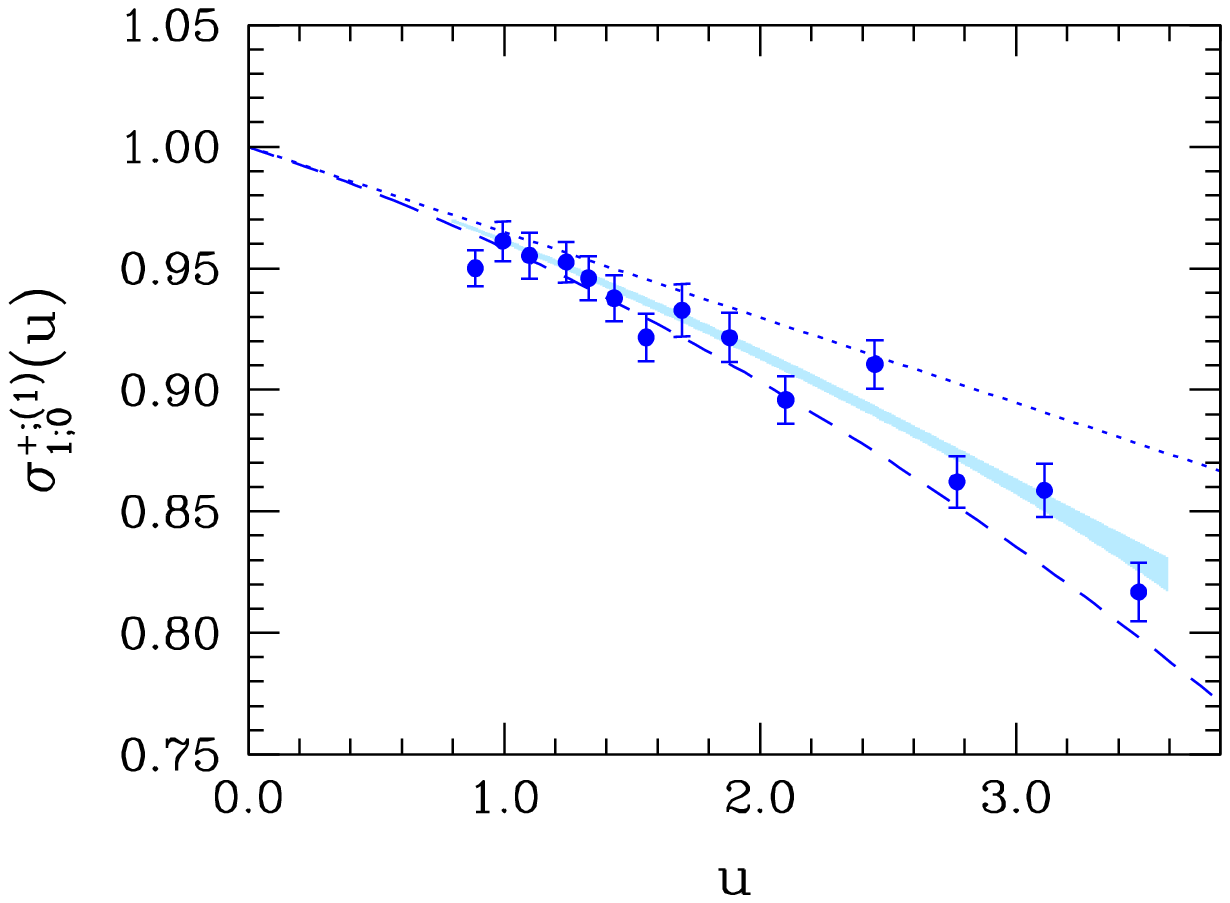}
\includegraphics{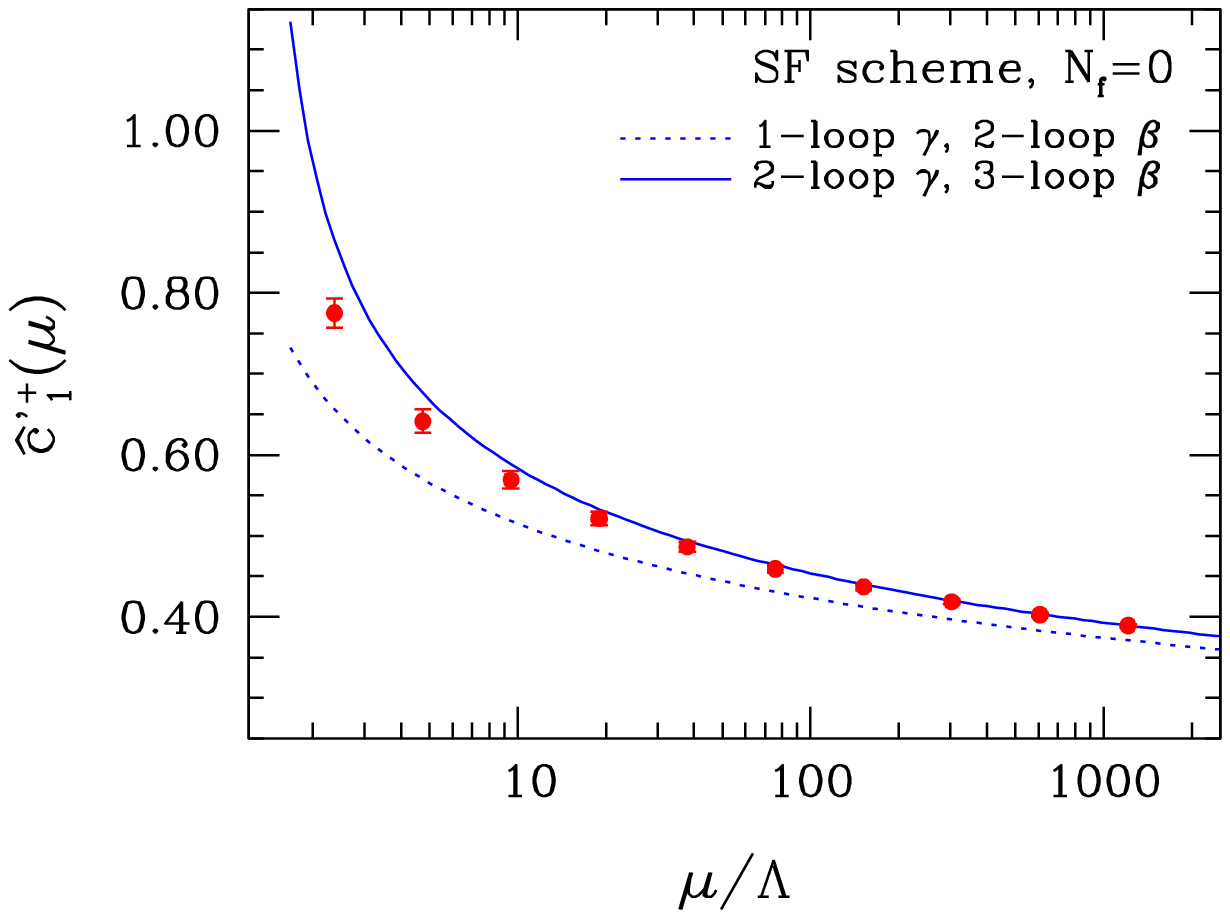}
\includegraphics{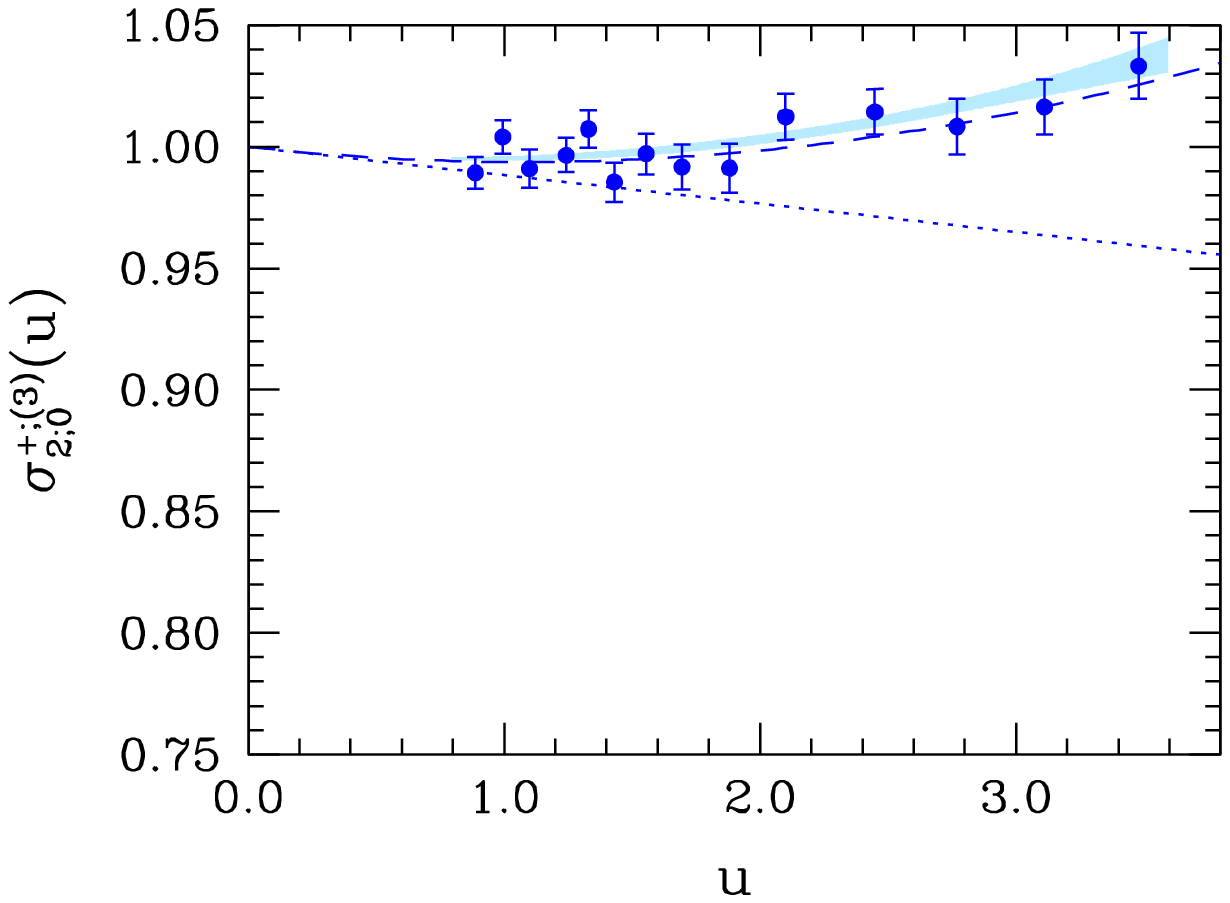}
\includegraphics{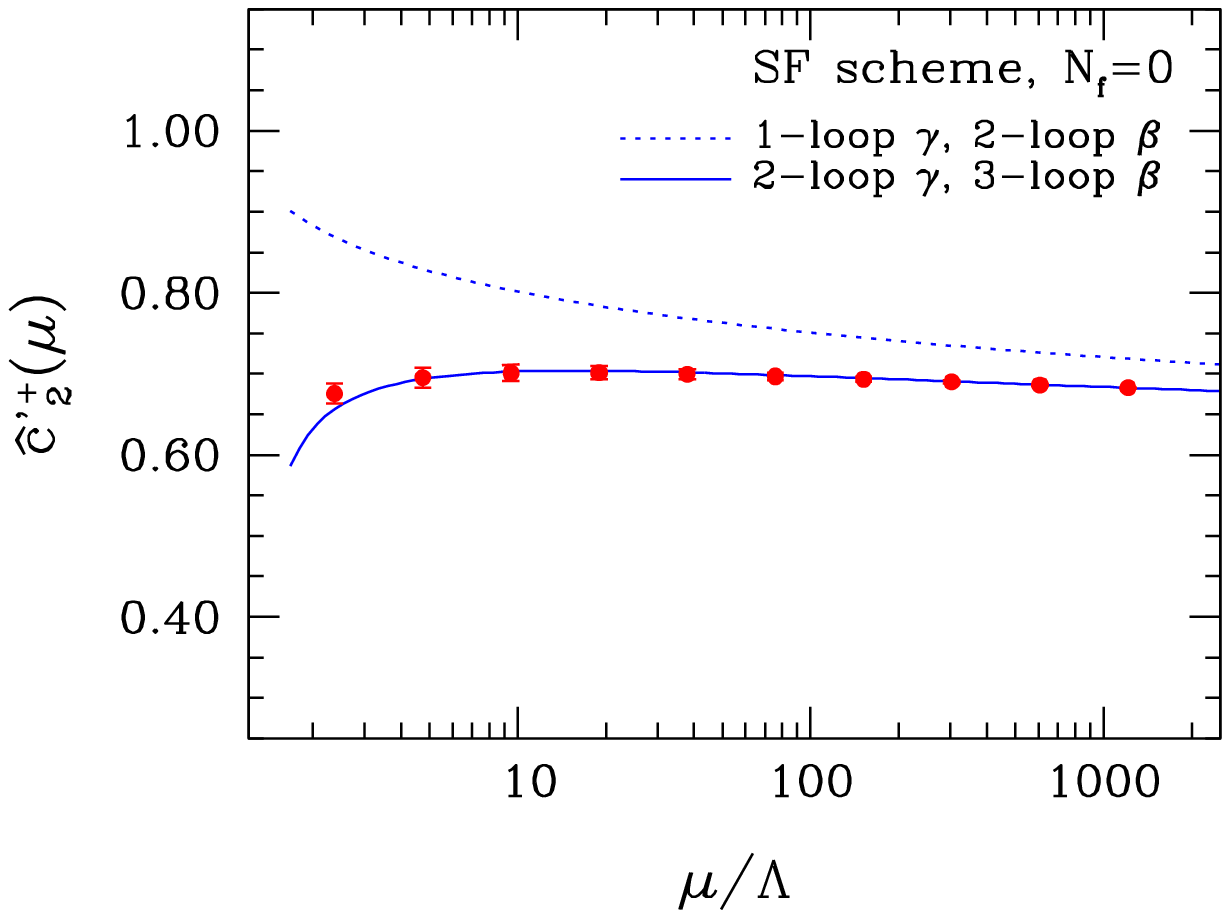}
\includegraphics{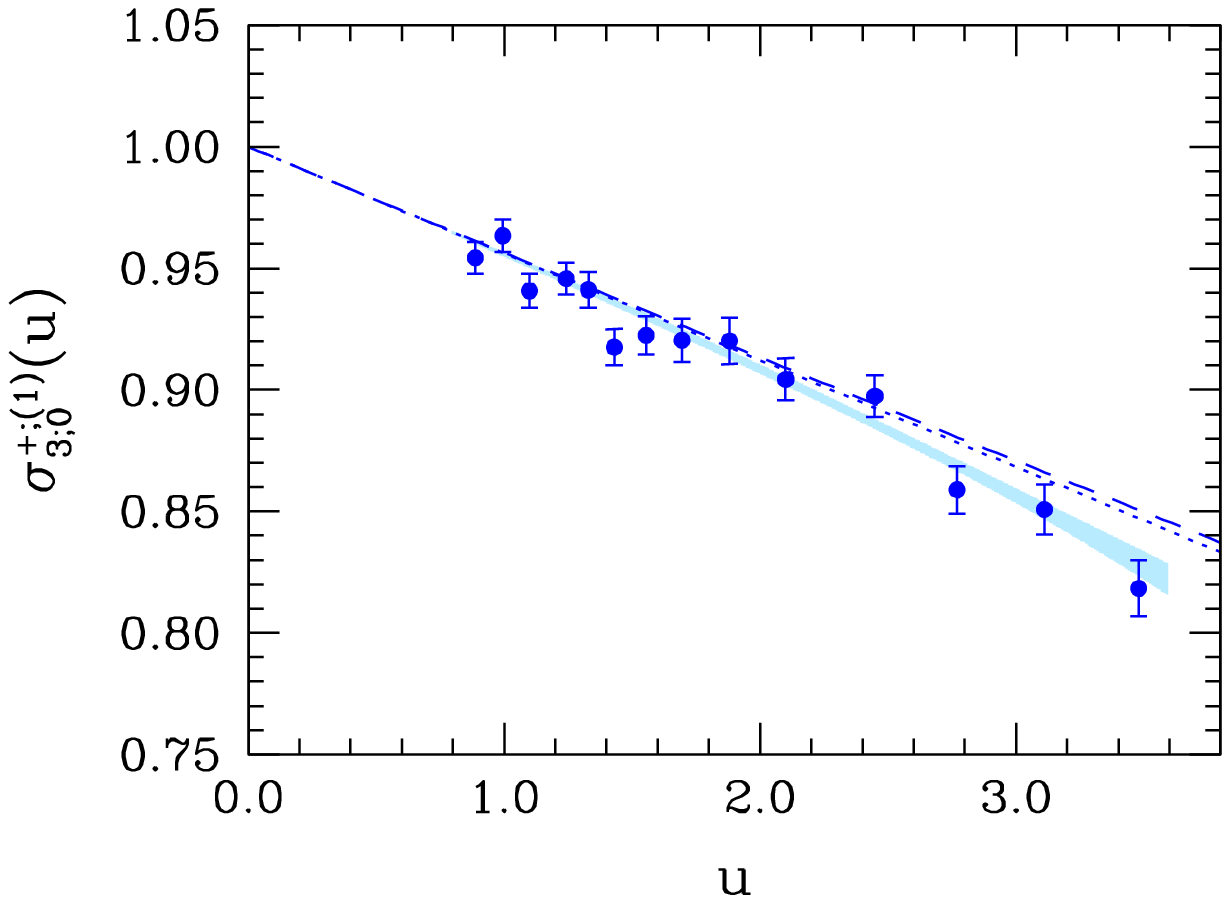}
\includegraphics{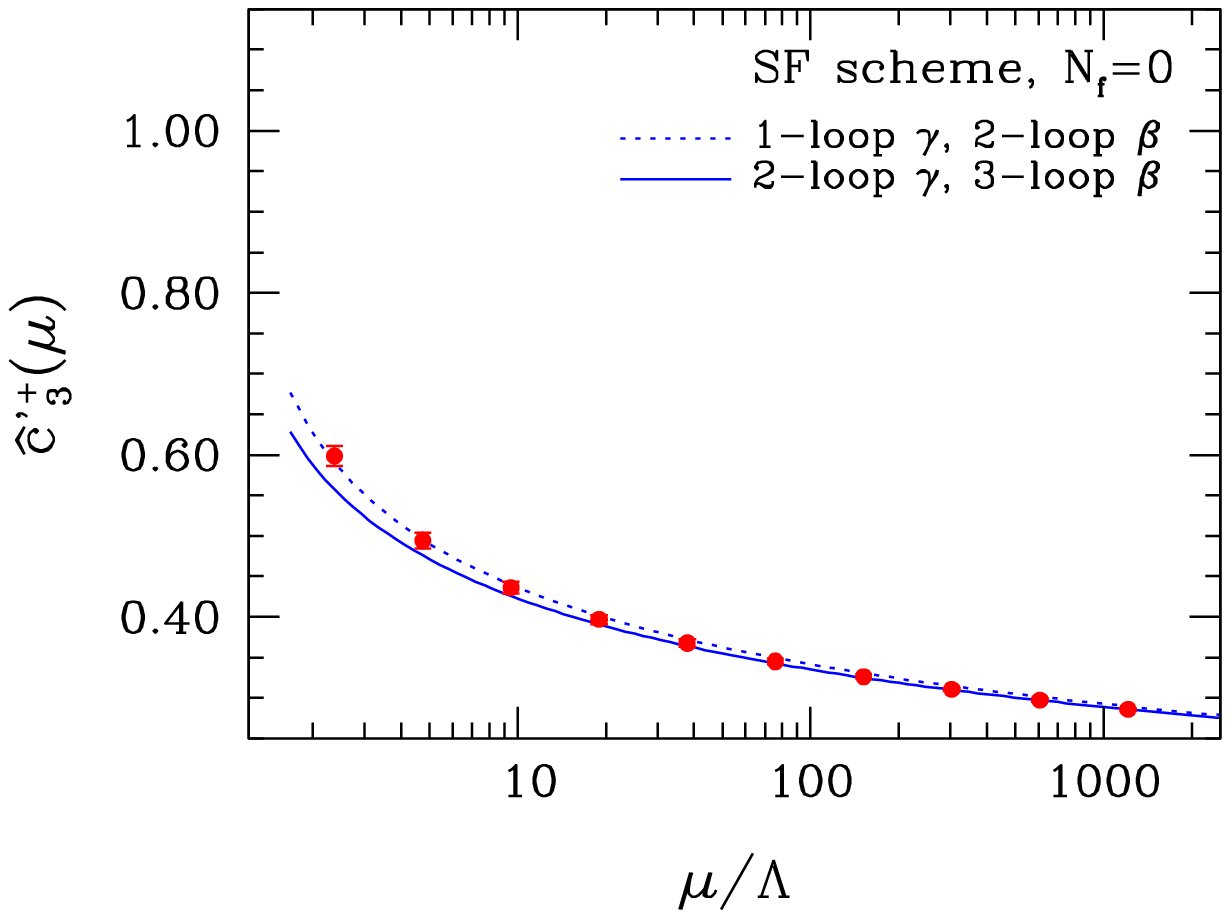}
\includegraphics{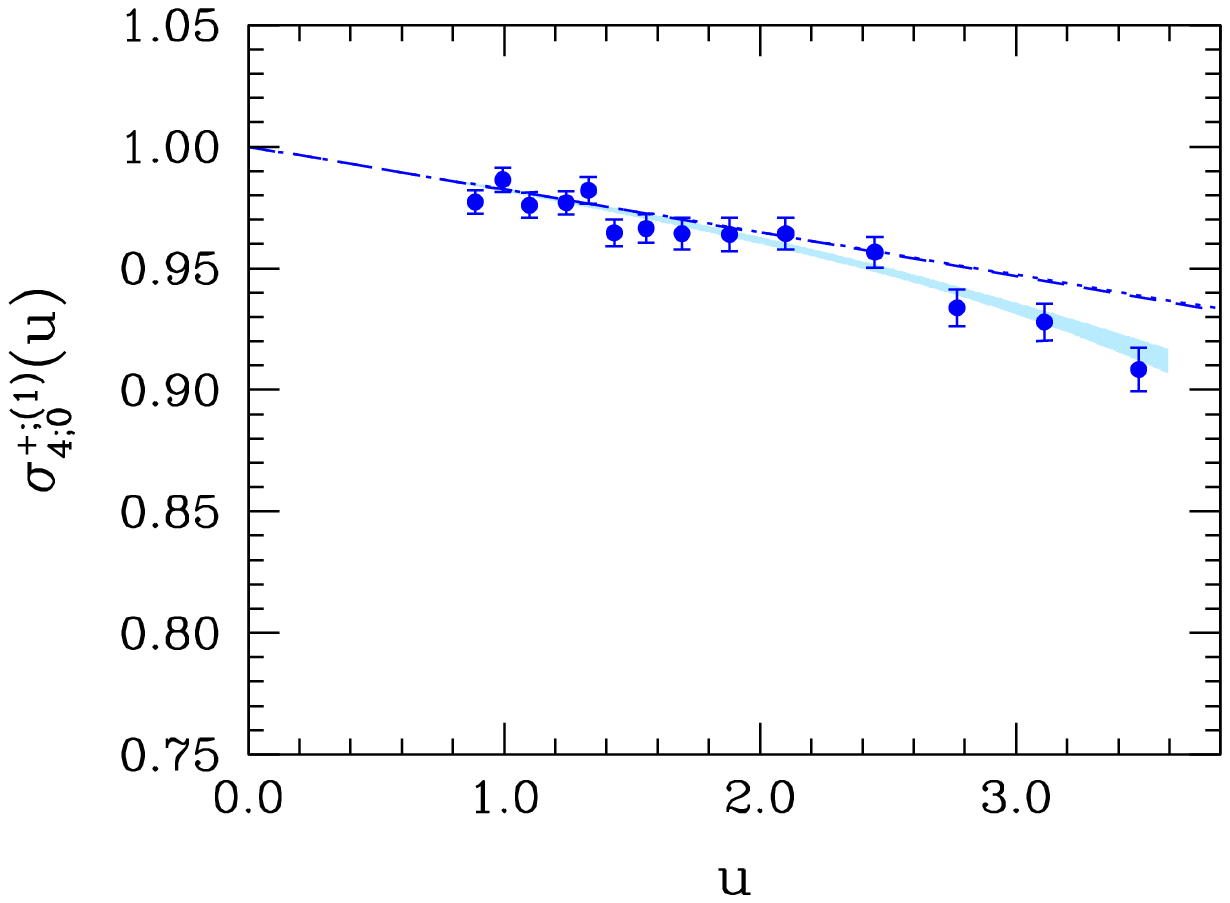}
\includegraphics{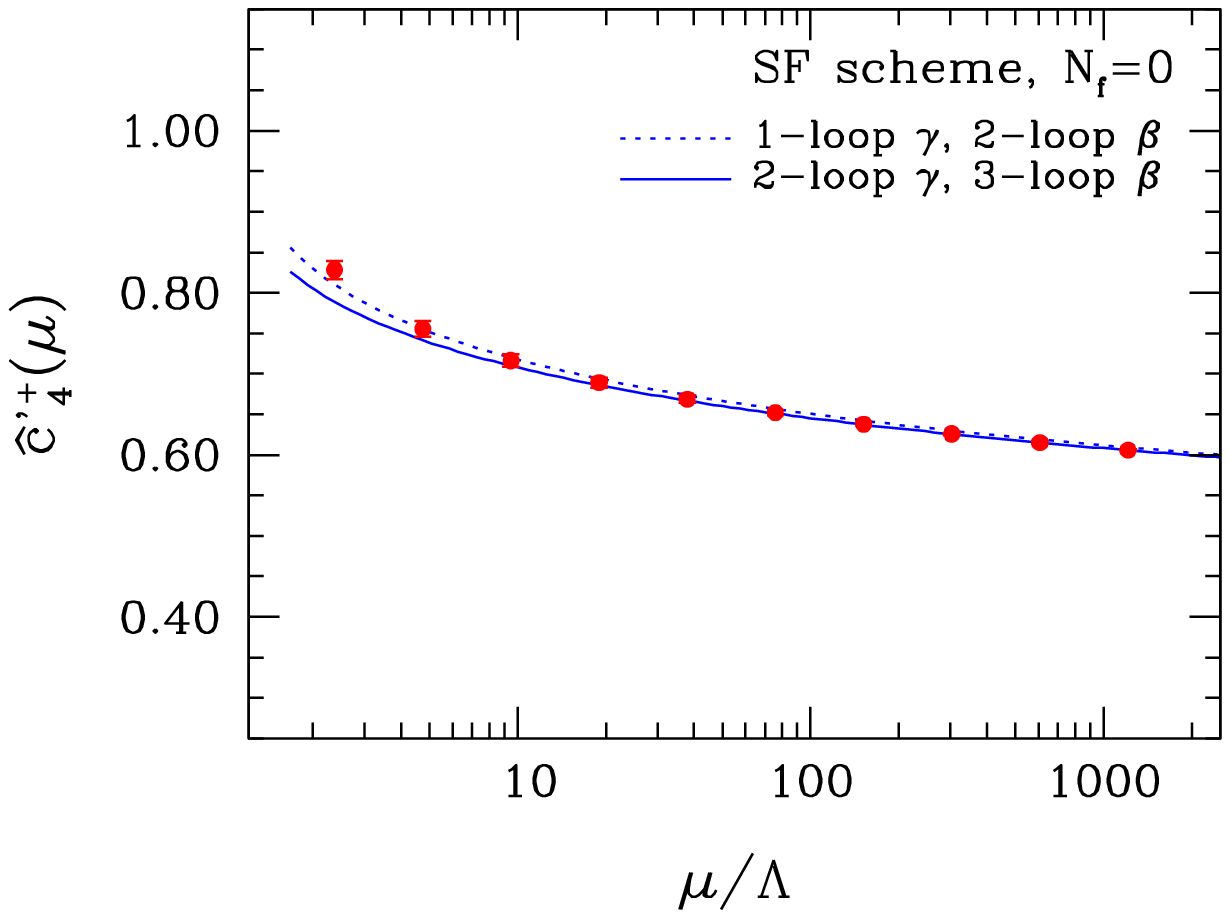}
\vspace{-75mm}
\caption{Left column: the step scaling function $\sigma_{k}^{\lp}$
(discrete points) as obtained non-perturbatively. The shaded area is the one sigma band
obtained by fitting the points to a polynomial. The dotted (dashed) line is the LO (NLO)
perturbative result. Right column: RG running of ${\mathcal{Q}'}_k^+$
obtained non perturbatively (discrete points) at specific
values of the renormalization scale $\mu$, in units of $\Lambda$.
The lines are pertrubative results at the order shown for the 
Callan-Symanzik $\beta$ function and the operator anomalous dimension
$\gamma$.
}
\label{fig:ssf_and_rg}
\end{figure}

The analysis described above yields accurate estimates of the continuum
SSFs $\sigma_{k}^{\lp}$ for a wide range of values of the renormalized
coupling. In order to compute the RG running of the operators in the
continuum limit as described in sect.~\ref{sec_lattice}, we need to
fit these data, as well as those for the SSF of the renormalized
coupling itself, to some functional form. Regarding the SSF of the coupling
$\sigma(u)$, we have followed the same procedure as
in~\cite{mbar:pap1}. This has been also adopted for the SSFs of the 
four-fermion operators, for which we have assumed the polynomial ansatz
\begin{gather}
\label{eq:ssf_ansatz}
\sigma_{k}^{\lp} = 1 + \sum_{n=1}^N s_n u^n \,,
\end{gather}
motivated by the form of the perturbative series. In particular, the
analytical expressions of the first two coefficients are given in perturbation 
theory by
\begin{gather}
\label{eq:pt_s1}
s_1 = {\gamma'}_k^{+;(0)}\ln 2 \,,\\
\label{eq:pt_s2}
s_2 = {\gamma'}_k^{+;(1)}\ln 2 + \left[
\frac{1}{2}({\gamma'}_k^{+;(0)})^2 + b_0{\gamma'}_k^{+;(0)}
\right](\ln 2)^2 \,.
\end{gather}
While the first coefficient is entirely determined by the LO anomalous
dimension and is hence universal, the second one, where the NLO
anomalous dimension enters, is scheme dependent. Contrary to the case
of the fully relativistic four-quark operators considered 
in~\cite{Palombi:2005zd}, the NLO coefficient does not depend strongly 
on the chosen SF scheme.
\parbreak
We have performed fits to the ansatz of Eq.~(\ref{eq:ssf_ansatz}) with
$N$ ranging from~$2$ to~$4$. The coefficient $s_1$ is always kept
fixed to the value in Eq.~(\ref{eq:pt_s1}), and fits are performed
either with $s_2$ fixed to the value in Eq.~(\ref{eq:pt_s2}) or
keeping it as a free parameter. All fits are well behaved, with values
of $\chi^2/\mbox{d.o.f.}$ ranging from~$0.8$ to~$1.6$. It is worth
mentioning that when $s_2$ is kept as a free parameter, its fitted
value lies in the ballpark of the perturbative prediction
of Eq.~(\ref{eq:pt_s2}), which can be taken as an indication that
perturbation theory indeed describes the data well within a large part
of the range of scales covered by our simulations. However, our data
are not accurate enough to allow for a more detailed check of the
applicability of perturbation theory beyond leading order.
\parbreak
Once a definite expression for the fitted step scaling function is
chosen, the solution of the recursion relations provided by 
Eqs.~(\ref{eq:coupevol}) and~(\ref{eq:NPevol}) is unique. At that
point, the value obtained for the RG running factor
${U'}_{k}^{\lp}(2^n\mu_{\rm had},\mu_{\rm had})$ of
Eq.~(\ref{eq:NPevol}) is a function of the fit parameters only. We have
checked that increasing the number of fit parameters provides
compatible results for ${U'}_{k}^{\lp}(2^n\mu_{\rm had},\mu_{\rm
had})$ with slightly larger errors. The result is also fairly
insensitive to whether $s_2$ is fixed to the perturbative prediction
or not. The conclusion is that, at the available level of precision,
the bias induced by the choice of the fit function is not significant,
which results in a numerically very stable determination of the
SSFs. We quote as our best results those coming from a two-parameter
fit with $s_1$ fixed by perturbation theory.
\parbreak
At this point it is useful to restrict the attention to a selected
subset of renormalization schemes. As discussed
in~\cite{Palombi:2006pu}, heavy quark spin symmetry provides a number
of identities between the 10 SF schemes we have considered per
operator\footnote{These identities have been verified explicitly
for each combination of the labels~$\alpha$ and~$s$ at all the
levels of our numerical analysis, which provides a check of the
latter.}. In practice, we have four different independent schemes for
the operators ${\cQ'}^\lp_1$ and ${\cQ'}^\lp_3$, and another eight for ${\cQ'}^\lp_2$
and ${\cQ'}^\lp_4$. All these schemes should lead to the same RGI
quantities, since the total renormalization factors (see below) differ only
by cutoff effects. This could be used potentially to improve continuum
limit extrapolations by combining various schemes. However, the strong
statistical correlation between the different renormalization factors
is likely to produce only a small gain in precision. Therefore, we
choose for each operator just one single representative
scheme. This strategy has been seen to be justified in the fully
relativistic case, i.e. in the computation of $B_K$~\cite{Dimopoulos:2006dm,Dimopoulos:2007cn}.
\parbreak
As discussed in~\cite{Guagnelli:2005zc}, the main criterion to define 
suitable schemes amounts to checking that the systematic uncertainty
related to truncating at NLO the perturbative matching at the scale $\mu_{\rm
pt}\equiv 2^n\mu_{\rm had}$ in Eq.~(\ref{factorization}) is well under
control. This in turn requires an estimate of the size of the NNLO
contribution to ${\hat{c}}'^{\lp}_{\:k}(\mu_{\rm pt})$. To this purpose
we have re-computed ${\hat{c}}'^{\lp}_{\:k}(\mu_{\rm pt})$ with two
different values of the NNLO anomalous dimensions ${\gamma'}^{+;(2)}_k$:
in the first case we set ${\gamma'}^{+;(2)}_k/{\gamma'}^{+;(1)}_k={\gamma'}^{+;(1)}_k/{\gamma'}^{+;(0)}_k$;
in the second case, we guess ${\gamma'}^{+;(2)}_k$ by performing a
one-parameter fit to the SSF with $s_1$ and $s_2$ fixed by perturbation
theory, and then equating the resulting value of $s_3$ to its perturbative
expression
\begin{gather}
\begin{split}
s_3 = {\gamma'}^{+;(2)}_k\ln 2 &+ \left[
{\gamma'}^{+;(0)}_k{\gamma'}^{+;(1)}_k + 2b_0{\gamma'}^{+;(1)}_k
+ b_1{\gamma'}^{+;(0)}_k
\right](\ln 2)^2 + \\[2.0ex] &+ \left[
\frac{1}{6}\left({\gamma'}^{+;(0)}_k\right)^3 + b_0 \left({\gamma'}^{+;(0)}_k\right)^2
+\frac{4}{3}b_0^2{\gamma'}^{+;(0)}_k
\right](\ln 2)^3 \,.
\end{split}
\end{gather}
For the operators ${\cQ'}^\lp_{1,3,4}$, we find that in either case the
central value of the combination
${\hat{c}}'^{\lp}_{\:k}(\mu_{\rm had})\equiv
{\hat{c}}'^\lp_{\:k}(\mu_{\rm pt}) {U'}^\lp_k(\mu_{\rm pt},\mu_{\rm had})$
changes by a small fraction of the statistical error, of the order
$0.1\sigma$--$0.3\sigma$. There is no systematic dependence on the
choice of boundary operators or normalization factors in the
renormalization condition. We thus conclude that this particular
uncertainty is well covered by the statistical one and choose
as our reference schemes those labeled by $(s,\alpha)=(1,0)$.
\parbreak
As for the operator ${\cQ'}^\lp_2$, which carries relatively large NLO
anomalous dimensions, the effect can be as large as $0.8\sigma$ with
$s=3$, and of the order of $1\sigma$ with the other values of $s$.
There is no significant dependence on $\alpha$. We therefore opt,
conservatively, for $(s,\alpha)=(3,0)$ as our preferred choice for this
operator, adding to ${\hat{c}}'^{\lp}_{\:2}(\mu_{\rm had})$ a systematic
uncertainty of~$0.8$ standard deviations. It has to be stressed that
the impact of this extra uncertainty at the level of the $B$--$\bar B$
mixing amplitude is not particularly worrying, since the matrix
element of ${\cQ'}^\lp_2$ enters the latter only at $\cO(\alpha_{\rm s})$ when
the static theory is matched to QCD.
It is therefore expected
to contribute a relatively small fraction to the final uncertainty.
\parbreak
The results for the operator RG running in these schemes are provided in 
Table~\ref{tab:fit_ztot1}. Those concerning the SSFs are collected in 
Table~\ref{tab:CLe1}. The same results are illustrated by 
Figure~\ref{fig:ssf_and_rg}.

\subsection{Matching to hadronic observables}

The RGI operator, defined in Eq.~(\ref{RGsol}), is connected to its
bare counterpart via a total renormalization factor
$\hat{Z}'^\lp_{\ k,\RGI}(g_0)$, as in Eq.\,(\ref{eq:totalren}). We stress
that $\hat{Z}'^\lp_{\ k,\RGI}(g_0)$ is a scale-independent quantity, which
moreover depends on the renormalization scheme only via cutoff
effects. Indeed, it depends on the particular lattice regularization
chosen, though only through the factor ${\cZ'}_{k}^{\lp}(g_0,a\mu_{\rm
had})$, the computation of which is much less expensive than the total RG 
running factor ${\hat{c}}'^{\lp}_{\:k}(\mu_{\rm had})$.
\parbreak
We have computed ${\cZ'}_{k}^{+}(g_0,a\mu_{\rm had}),
\mu_{\rm{had}}=1/(2L_{\rm max})$ non-perturbatively at four values of
$\beta$ for each scheme and four-fermion operator, and for the four
different static actions under consideration. The results for the HYP2
action and the reference renormalization schemes defined in
sect.~\ref{sec:RGrunning} are given in Table~\ref{tab:Zm1}. Upon
multiplying by the corresponding running factors in Table~\ref{tab:fit_ztot1}, the
total renormalization factors are obtained. These can be further
fitted to polynomials of the form
\begin{gather}
  \hat{Z}'^\lp_{\ k,\RGI}(g_0)
    =a_{k} +b_{k}(\beta-6) +c_{k}(\beta-6)^2\,,
  \label{eq:Ztotfit}
\end{gather}
which can be subsequently used to obtain the total renormalization
factor at any value of $\beta$ within the covered range.\footnote{Note
that $\beta=6.0$ lies very slightly below the interval covered
by our simulations, and therefore a very small extrapolation
is required.}
We provide in Table~\ref{tab:fit_ztot1} the resulting fit
coefficients for the HYP2 action in our reference renormalization
schemes. These parameterizations represent our data with an accuracy
of at least $0.3\%$ (this comprises the point
$\beta=6.0$. The contribution from the error in the RG running
factors of Table~\ref{tab:fit_ztot1} has been neglected: since these factors
have been computed in the continuum limit, they should be added
in quadrature {\it after} the quantity renormalized with the
factor derived from Eq.~(\ref{eq:Ztotfit}) has been extrapolated itself
to the continuum limit.\newline

\TABLE[!t]{
\begin{tabular}{lc@{\hspace{10mm}}clll}
\hline\\[-1.0ex]
$k$ & $s$ & ${\hat{c}}'^{\lp;(s)}_{\:k;0}(\mu_{\rm had})$ & $~~~~a_{k;0}^{(s)}$ & $~~~~b_{k;0}^{(s)}$ & $~~~c_{k;0}^{(s)}$ \\[1.0ex]
\hline\\[-1.0ex]
$1$   & $1$ & $0.777(17)$ & $0.5731(11)$ & $-0.171(11)$ & $0.082(25)$ \\
$2^*$ & $3$ & $0.675(12)$ & $0.7258(14)$ & $-0.061(14)$ & $0.016(33)$ \\
$3$   & $1$ & $0.598(12)$ & $0.4564(8)$  & $-0.142(8)$  & $0.068(18)$ \\
$4$   & $1$ & $0.828(11)$ & $0.6465(9)$  & $-0.070(9)$  & $0.033(21)$ \\[1.0ex]
\hline
\end{tabular}
\caption{
Running factor ${\hat{c}}'^{\lp;(s)}_{\:k;0}(\mu_{\rm had})$ and fit coefficients (see Eq.~(\ref{eq:Ztotfit}))
to the total renormalization factor $\hat{Z}'^\lp_{\ k,\RGI}(g_0)$ introduced in Eq.~(\ref{eq:totalren}). 
Here $\mu_{\rm had}^{-1}=2L_{\rm max}$. The schemes characterized by larger systematic uncertainties related 
to perturbation theory have been indicated  with an asterisk.}
\label{tab:fit_ztot1}
}

\section{Conclusions \label{sec_concl}}

$B^0-\bar{B}^0$ mixing remains among the most important processes that
are required to pin down the elements of the CKM
matrix precisely. However, in order to constrain the unitarity triangle
sufficiently well and to look for signs of new physics, theoretical
uncertainties associated with hadronic effects must be further
reduced. Non-perturbative renormalization of four-quark operators is
an indispensable ingredient to enable lattice determinations of the
corresponding hadronic matrix elements with a total accuracy at the
level of a few percent.
\parbreak
In this paper we have described our fully non-perturbative calculation
of the relations between parity-odd, static-light four-quark operators
in quenched lattice QCD and their renormalized counterparts. Our main
results for the complete basis of operators are expressed by the
interpolating formula of Eq.~(\ref{eq:Ztotfit}), in conjunction with
the coefficients listed in Table~\ref{tab:fit_ztot1}. In addition to
the regularization-dependent, total renormalization factors
$\hat{Z'}^\lp_{k,\RGI}$, we also list the universal running factors
${\hat{c}}'^{\lp}_{\:k}(\mu_{\rm had})$, which, if desired, can be
combined with a different fermionic discretization, provided that the
regularization-dependent matching factor ${\cZ'}_k^\lp(g_0,a\mu_{\rm
had})$ (c.f. Eq.~(\ref{eq:totalren})) is re-computed.
\parbreak
The bulk of the uncertainty associated with the renormalization
originates from the universal running factors, which have been
determined with an accuracy of $1.5-2$\%. The level of precision
ensures that the targeted accuracy of, say, 5\% in the final result
for the $B$-parameters in the continuum limit can be reached. The
calculation of the bare hadronic matrix elements in quenched twisted
mass QCD is currently underway.
\parbreak
Finally, we stress that our method can be straightforwardly extended
to the unquenched case, with the simulations to compute the
step scaling functions for $\Nf=2$ dynamical quark flavours currently
in progress \cite{4fstatnf2}. Although Ginsparg-Wilson fermions appear as the natural
discretization to study left-left four-quark operators, the
Schr\"odinger functional is far more complicated to implement than for
Wilson-like fermions. In our approach, tmQCD serves to solve the
intricate renormalization problem for four-quark operators, while the
SF scheme is easy to implement and dynamical simulations with Wilson
fermions can be performed in an economical way.

\section*{Acknowledgements}
We thank R.~Sommer for useful discussions, and J.~Heitger
for his kind technical help. F.P.
acknowledges financial support from the
Alexander-von-Humboldt Stiftung. M.P. acknowledges financial support 
by an EIF Marie Curie fellowship of the European Community's Sixth 
Framework Programme under contract number MEIF-CT-2006-040458.
C.P. acknowledges partial financial support
by CICyT project FPA2006-05807.
We also thank the DESY Zeuthen computing centre 
for technical support. This work was supported in part 
by the EU Contract No. MRTN-CT-2006-035482, ``FLAVIAnet''.

\appendix

\newpage

\section[Appendix A]{Tables and figures\label{app:A}}

\TABLE[!h]{
\footnotesize
\centering
\begin{tabular}{rrllccc}
\hline \\[-1.0ex]
$~~~~~~~\beta~~~~$ & ${L}/{a}$ & $~~~\gbar^2_{\rm\scriptscriptstyle SF}(L)$ &
$~~~~~\hopc$ & $\mathcal{Z'}_{1;0}^{+;(1)}\left(g_0,{a}/{L}\right)$ & $\mathcal{Z'}_{1;0}^{+;(1)}\left(g_0,{a}/{2L}\right)$ &\
 ${\Sigma}^{+;(1)}_{1;0}\left(g_0,{a}/{L}\right)$ \\[1.0ex]
\hline \\[-1.0ex]
10.7503 & 6 & 0.8873(5) & 0.130591(4) & 0.9136(8) & 0.8827(12) & 0.9662(15) \\ 
11.0000 & 8 & 0.8873(10) & 0.130439(3) & 0.9041(7) & 0.8751(24) & 0.9679(28) \\ 
11.3384 & 12 & 0.8873(30) & 0.130251(2) & 0.8912(7) & 0.8571(26) & 0.9617(30) \\ 
11.5736 & 16 & 0.8873(25) & 0.130125(2) & 0.8827(14) & 0.8467(35) & 0.9592(43) \\ 
  [1.0ex]
  \hline\\[-1.0ex]
10.0500 & 6 & 0.9944(7) & 0.131073(5) & 0.9073(8) & 0.8714(8) & 0.9604(12) \\ 
10.3000 & 8 & 0.9944(13) & 0.130889(3) & 0.8943(8) & 0.8590(28) & 0.9605(32) \\ 
10.6086 & 12 & 0.9944(30) & 0.130692(2) & 0.8798(7) & 0.8478(25) & 0.9636(30) \\ 
10.8910 & 16 & 0.9944(28) & 0.130515(2) & 0.8737(25) & 0.8365(34) & 0.9574(47) \\ 
  [1.0ex]
  \hline\\[-1.0ex]
9.5030 & 6 & 1.0989(8) & 0.131514(5) & 0.8992(9) & 0.8611(11) & 0.9576(15) \\ 
9.7500 & 8 & 1.0989(13) & 0.131312(3) & 0.8876(8) & 0.8484(33) & 0.9558(38) \\ 
10.0577 & 12 & 1.0989(40) & 0.131079(3) & 0.8713(10) & 0.8311(32) & 0.9539(38) \\ 
10.3419 & 16 & 1.0989(44) & 0.130876(2) & 0.8645(33) & 0.8274(33) & 0.9571(53) \\ 
  [1.0ex]
  \hline\\[-1.0ex]
8.8997 & 6 & 1.2430(13) & 0.132072(9) & 0.8894(9) & 0.8459(12) & 0.9511(17) \\ 
9.1544 & 8 & 1.2430(14) & 0.131838(4) & 0.8781(9) & 0.8314(34) & 0.9468(40) \\ 
9.5202 & 12 & 1.2430(35) & 0.131503(3) & 0.8613(8) & 0.8177(23) & 0.9494(28) \\ 
9.7350 & 16 & 1.2430(34) & 0.131335(3) & 0.8490(19) & 0.8058(31) & 0.9491(42) \\ 
  [1.0ex]
  \hline\\[-1.0ex]
8.6129 & 6 & 1.3293(12) & 0.132380(6) & 0.8854(10) & 0.8391(12) & 0.9477(17) \\ 
8.8500 & 8 & 1.3293(21) & 0.132140(5) & 0.8714(9) & 0.8192(41) & 0.9401(48) \\ 
9.1859 & 12 & 1.3293(60) & 0.131814(3) & 0.8545(12) & 0.8069(35) & 0.9443(43) \\ 
9.4381 & 16 & 1.3293(40) & 0.131589(2) & 0.8400(18) & 0.7915(30) & 0.9423(41) \\ 
  [1.0ex]
  \hline\\[-1.0ex]
8.3124 & 6 & 1.4300(20) & 0.132734(10) & 0.8810(10) & 0.8308(12) & 0.9430(17) \\ 
8.5598 & 8 & 1.4300(21) & 0.132453(5) & 0.8668(10) & 0.8104(39) & 0.9349(46) \\ 
8.9003 & 12 & 1.4300(50) & 0.132095(3) & 0.8474(9) & 0.7947(40) & 0.9378(48) \\ 
9.1415 & 16 & 1.4300(58) & 0.131855(3) & 0.8304(18) & 0.7770(31) & 0.9357(43) \\ 
  [1.0ex]
  \hline\\[-1.0ex]
7.9993 & 6 & 1.5553(15) & 0.133118(7) & 0.8725(10) & 0.8126(14) & 0.9313(20) \\ 
8.2500 & 8 & 1.5553(24) & 0.132821(5) & 0.8573(11) & 0.8051(40) & 0.9391(48) \\ 
8.5985 & 12 & 1.5533(70) & 0.132427(3) & 0.8380(20) & 0.7850(39) & 0.9368(52) \\ 
8.8323 & 16 & 1.5533(70) & 0.132169(3) & 0.8261(19) & 0.7677(32) & 0.9293(44) \\ 
[1.0ex]
  \hline\\[-1.0ex]
\end{tabular}
\caption{
Numerical values of the renormalization constant $\mathcal{Z'}_{1;0}^{+;(1)}$ 
and the step scaling function ${\Sigma}_{1;0}^{+;(1)}$ with HYP2 action at various
renormalized SF couplings and lattice spacings. 
}
\label{tab:tab3}
}

\vfill\eject

\vbox{\ }
\vskip 0.48cm

\TABLE[!h]{
\footnotesize
\centering
\begin{tabular}{rrllccc}
\hline \\[-1.0ex]
$~~~~~~~\beta~~~~$ & ${L}/{a}$ & $~~~\gbar^2_{\rm\scriptscriptstyle SF}(L)$ &
$~~~~~\hopc$ & $\mathcal{Z'}_{1;0}^{+;(1)}\left(g_0,{a}/{L}\right)$ & $\mathcal{Z'}_{1;0}^{+;(1)}\left(g_0,{a}/{2L}\right)$ &\
 ${\Sigma}_{1;0}^{+;(1)}\left(g_0,{a}/{L}\right)$ \\[1.0ex]
\hline \\[-1.0ex]
7.7170 & 6 & 1.6950(26) & 0.133517(8) & 0.8664(11) & 0.8024(10) & 0.9261(17) \\ 
7.9741 & 8 & 1.6950(28) & 0.133179(5) & 0.8503(9) & 0.7836(40) & 0.9216(48) \\ 
8.3218 & 12 & 1.6950(79) & 0.132756(4) & 0.8292(10) & 0.7684(37) & 0.9267(46) \\ 
8.5479 & 16 & 1.6950(90) & 0.132485(3) & 0.8154(21) & 0.7553(41) & 0.9263(55) \\ 
  [1.0ex]
  \hline\\[-1.0ex]
7.4082 & 6 & 1.8811(22) & 0.133961(8) & 0.8571(12) & 0.7891(24) & 0.9207(30) \\ 
7.6547 & 8 & 1.8811(28) & 0.133632(6) & 0.8399(11) & 0.7705(38) & 0.9174(47) \\ 
7.9993 & 12 & 1.8811(38) & 0.133159(4) & 0.8192(15) & 0.7493(35) & 0.9147(46) \\ 
8.2415 & 16 & 1.8811(99) & 0.132847(3) & 0.7983(28) & 0.7356(30) & 0.9215(50) \\ 
  [1.0ex]
  \hline\\[-1.0ex]
7.1214 & 6 & 2.1000(39) & 0.134423(9) & 0.8478(13) & 0.7687(17) & 0.9067(24) \\ 
7.3632 & 8 & 2.1000(45) & 0.134088(6) & 0.8304(12) & 0.7512(33) & 0.9046(42) \\ 
7.6985 & 12 & 2.1000(80) & 0.133599(4) & 0.8040(15) & 0.7241(34) & 0.9006(46) \\ 
7.9560 & 16 & 2.100(11) & 0.133229(3) & 0.7889(21) & 0.7107(36) & 0.9009(51) \\ 
  [1.0ex]
  \hline\\[-1.0ex]
6.7807 & 6 & 2.4484(37) & 0.134994(11) & 0.8339(21) & 0.7414(22) & 0.8891(35) \\ 
7.0197 & 8 & 2.4484(45) & 0.134639(7) & 0.8124(13) & 0.7173(28) & 0.8829(38) \\ 
7.3551 & 12 & 2.4484(80) & 0.134141(5) & 0.7867(19) & 0.7079(34) & 0.8998(48) \\ 
7.6101 & 16 & 2.448(17) & 0.133729(4) & 0.7734(25) & 0.6899(34) & 0.8920(53) \\ 
  [1.0ex]
  \hline\\[-1.0ex]
6.5512 & 6 & 2.770(7) & 0.135327(12) & 0.8245(15) & 0.7146(20) & 0.8667(29) \\ 
6.7860 & 8 & 2.770(7) & 0.135056(8) & 0.7995(14) & 0.6946(36) & 0.8688(48) \\ 
7.1190 & 12 & 2.770(11) & 0.134513(5) & 0.7731(14) & 0.6700(40) & 0.8666(54) \\ 
7.3686 & 16 & 2.770(14) & 0.134114(3) & 0.7546(24) & 0.6531(34) & 0.8655(53) \\ 
  [1.0ex]
  \hline\\[-1.0ex]
6.3665 & 6 & 3.111(4) & 0.135488(6) & 0.8166(15) & 0.6884(20) & 0.8430(29) \\ 
6.6100 & 8 & 3.111(6) & 0.135339(3) & 0.7902(14) & 0.6688(32) & 0.8464(43) \\ 
6.9322 & 12 & 3.111(12) & 0.134855(3) & 0.7585(18) & 0.6487(42) & 0.8552(59) \\ 
7.1911 & 16 & 3.111(16) & 0.134411(3) & 0.7399(24) & 0.6291(36) & 0.8503(56) \\ 
  [1.0ex]
  \hline\\[-1.0ex]
6.2204 & 6 & 3.480(8) & 0.135470(15) & 0.8062(16) & 0.6574(21) & 0.8154(30) \\ 
6.4527 & 8 & 3.480(14) & 0.135543(9) & 0.7791(14) & 0.6458(44) & 0.8289(59) \\ 
6.7750 & 12 & 3.480(39) & 0.135121(5) & 0.7452(18) & 0.6162(30) & 0.8269(46) \\ 
7.0203 & 16 & 3.480(21) & 0.134707(4) & 0.7243(22) & 0.5951(39) & 0.8216(60) \\ 
  [1.0ex]
\hline\\[-1.0ex]
\end{tabular}
\caption{
Numerical values of the renormalization constant $\mathcal{Z'}_{1;0}^{+;(1)}$ 
and the step scaling function ${\Sigma}_{1;0}^{+;(1)}$ with HYP2 action at various
renormalized SF couplings and lattice spacings (continued).}
\label{tab:tab4}
}

\vfill\eject

\vbox{\ }
\vskip 0.48cm

\TABLE[!h]{
\footnotesize
\centering
\begin{tabular}{rrllccc}
\hline \\[-1.0ex]
$~~~~~~~\beta~~~~$ & ${L}/{a}$ & $~~~\gbar^2_{\rm\scriptscriptstyle SF}(L)$ &
$~~~~~\hopc$ & $\mathcal{Z'}_{2;0}^{+;(3)}\left(g_0,{a}/{L}\right)$ & $\mathcal{Z'}_{2;0}^{+;(3)}\left(g_0,{a}/{2L}\right)$ &\
 ${\Sigma}_{2;0}^{+;(3)}\left(g_0,{a}/{L}\right)$ \\[1.0ex]
\hline \\[-1.0ex]
10.7503 & 6 & 0.8873(5) & 0.130591(4) & 1.0020(7) & 0.9905(11) & 0.9885(13) \\ 
11.0000 & 8 & 0.8873(10) & 0.130439(3) & 0.9936(7) & 0.9847(24) & 0.9910(25) \\ 
11.3384 & 12 & 0.8873(30) & 0.130251(2) & 0.9885(6) & 0.9798(24) & 0.9912(25) \\ 
11.5736 & 16 & 0.8873(25) & 0.130125(2) & 0.9851(13) & 0.9744(36) & 0.9891(39) \\ 
  [1.0ex]
  \hline\\[-1.0ex]
10.0500 & 6 & 0.9944(7) & 0.131073(5) & 1.0033(8) & 0.9897(8) & 0.9864(11) \\ 
10.3000 & 8 & 0.9944(13) & 0.130889(3) & 0.9956(7) & 0.9838(26) & 0.9881(27) \\ 
10.6086 & 12 & 0.9944(30) & 0.130692(2) & 0.9881(6) & 0.9825(24) & 0.9943(25) \\ 
10.8910 & 16 & 0.9944(28) & 0.130515(2) & 0.9835(23) & 0.9786(31) & 0.9950(39) \\ 
  [1.0ex]
  \hline\\[-1.0ex]
9.5030 & 6 & 1.0989(8) & 0.131514(5) & 1.0040(9) & 0.9886(10) & 0.9847(13) \\ 
9.7500 & 8 & 1.0989(13) & 0.131312(3) & 0.9966(8) & 0.9851(30) & 0.9885(31) \\ 
10.0577 & 12 & 1.0989(40) & 0.131079(3) & 0.9891(10) & 0.9806(27) & 0.9914(29) \\ 
10.3419 & 16 & 1.0989(44) & 0.130876(2) & 0.9876(30) & 0.9752(34) & 0.9874(45) \\ 
  [1.0ex]
  \hline\\[-1.0ex]
8.8997 & 6 & 1.2430(13) & 0.132072(9) & 1.0061(10) & 0.9888(12) & 0.9828(15) \\ 
9.1544 & 8 & 1.2430(14) & 0.131838(4) & 0.9978(9) & 0.9858(31) & 0.9880(33) \\ 
9.5202 & 12 & 1.2430(35) & 0.131503(3) & 0.9889(8) & 0.9807(23) & 0.9917(24) \\ 
9.7350 & 16 & 1.2430(34) & 0.131335(3) & 0.9835(18) & 0.9751(32) & 0.9915(37) \\ 
  [1.0ex]
  \hline\\[-1.0ex]
8.6129 & 6 & 1.3293(12) & 0.132380(6) & 1.0095(11) & 0.9918(13) & 0.9825(16) \\ 
8.8500 & 8 & 1.3293(21) & 0.132140(5) & 0.9994(10) & 0.9860(39) & 0.9866(40) \\ 
9.1859 & 12 & 1.3293(60) & 0.131814(3) & 0.9898(12) & 0.9794(30) & 0.9895(33) \\ 
9.4381 & 16 & 1.3293(40) & 0.131589(2) & 0.9828(17) & 0.9813(30) & 0.9985(35) \\ 
  [1.0ex]
  \hline\\[-1.0ex]
8.3124 & 6 & 1.4300(20) & 0.132734(10) & 1.0101(11) & 0.9921(12) & 0.9822(16) \\ 
8.5598 & 8 & 1.4300(21) & 0.132453(5) & 0.9986(10) & 0.9927(45) & 0.9941(46) \\ 
8.9003 & 12 & 1.4300(50) & 0.132095(3) & 0.9926(9) & 0.9724(36) & 0.9796(37) \\ 
9.1415 & 16 & 1.4300(58) & 0.131855(3) & 0.9864(20) & 0.9790(28) & 0.9925(35) \\ 
  [1.0ex]
  \hline\\[-1.0ex]
7.9993 & 6 & 1.5553(15) & 0.133118(7) & 1.0139(12) & 0.9909(16) & 0.9773(20) \\ 
8.2500 & 8 & 1.5553(24) & 0.132821(5) & 1.0015(12) & 0.9926(41) & 0.9911(43) \\ 
8.5985 & 12 & 1.5533(70) & 0.132427(3) & 0.9915(20) & 0.9855(43) & 0.9939(47) \\ 
8.8323 & 16 & 1.5533(70) & 0.132169(3) & 0.9844(19) & 0.9784(33) & 0.9939(38) \\ 
  [1.0ex]
\hline\\[-1.0ex]
\end{tabular}
\caption{
Numerical values of the renormalization constant $\mathcal{Z'}_{2;0}^{+;(3)}$ 
and the step scaling function ${\Sigma}_{2;0}^{+;(3)}$ with HYP2 action at various
renormalized SF couplings and lattice spacings.}
\label{tab:tab5}
}

\vfill\eject

\vbox{\ }
\vskip 0.48cm

\TABLE[!h]{
\footnotesize
\centering
\begin{tabular}{rrllccc}
\hline \\[-1.0ex]
$~~~~~~~\beta~~~~$ & ${L}/{a}$ & $~~~\gbar^2_{\rm\scriptscriptstyle SF}(L)$ &
$~~~~~\hopc$ & $\mathcal{Z'}_{2;0}^{+;(3)}\left(g_0,{a}/{L}\right)$ & $\mathcal{Z'}_{2;0}^{+;(3)}\left(g_0,{a}/{2L}\right)$ &\
 ${\Sigma}_{2;0}^{+;(3)}\left(g_0,{a}/{L}\right)$ \\[1.0ex]
\hline \\[-1.0ex]
7.7170 & 6 & 1.6950(26) & 0.133517(8) & 1.0153(13) & 0.9938(12) & 0.9788(17) \\ 
7.9741 & 8 & 1.6950(28) & 0.133179(5) & 1.0025(10) & 0.9892(40) & 0.9867(41) \\ 
8.3218 & 12 & 1.6950(79) & 0.132756(4) & 0.9928(11) & 0.9866(37) & 0.9938(39) \\ 
8.5479 & 16 & 1.6950(90) & 0.132485(3) & 0.9897(21) & 0.9753(43) & 0.9855(49) \\ 
  [1.0ex]
  \hline\\[-1.0ex]
7.4082 & 6 & 1.8811(22) & 0.133961(8) & 1.0137(13) & 1.0021(26) & 0.9886(28) \\ 
7.6547 & 8 & 1.8811(28) & 0.133632(6) & 1.0051(13) & 0.9927(56) & 0.9877(57) \\ 
7.9993 & 12 & 1.8811(38) & 0.133159(4) & 0.9917(16) & 0.9835(34) & 0.9917(38) \\ 
8.2415 & 16 & 1.8811(99) & 0.132847(3) & 0.9876(30) & 0.9762(31) & 0.9885(44) \\ 
  [1.0ex]
  \hline\\[-1.0ex]
7.1214 & 6 & 2.1000(39) & 0.134423(9) & 1.0249(15) & 1.0011(21) & 0.9768(25) \\ 
7.3632 & 8 & 2.1000(45) & 0.134088(6) & 1.0113(14) & 0.9971(43) & 0.9860(45) \\ 
7.6985 & 12 & 2.1000(80) & 0.133599(4) & 0.9944(17) & 0.9911(43) & 0.9967(46) \\ 
7.9560 & 16 & 2.100(11) & 0.133229(3) & 0.9857(22) & 0.9841(39) & 0.9984(46) \\ 
  [1.0ex]
  \hline\\[-1.0ex]
6.7807 & 6 & 2.4484(37) & 0.134994(11) & 1.0276(25) & 1.0153(30) & 0.9880(38) \\ 
7.0197 & 8 & 2.4484(45) & 0.134639(7) & 1.0158(15) & 1.0013(35) & 0.9857(37) \\ 
7.3551 & 12 & 2.4484(80) & 0.134141(5) & 1.0005(21) & 1.0035(43) & 1.0030(48) \\ 
7.6101 & 16 & 2.448(17) & 0.133729(4) & 0.9965(30) & 0.9923(39) & 0.9958(49) \\ 
  [1.0ex]
  \hline\\[-1.0ex]
6.5512 & 6 & 2.770(7) & 0.135327(12) & 1.0371(19) & 1.0174(27) & 0.9810(31) \\ 
6.7860 & 8 & 2.770(7) & 0.135056(8) & 1.0233(17) & 1.0144(55) & 0.9913(56) \\ 
7.1190 & 12 & 2.770(11) & 0.134513(5) & 1.0054(17) & 0.9972(54) & 0.9918(56) \\ 
7.3686 & 16 & 2.770(14) & 0.134114(3) & 0.9959(30) & 0.9976(45) & 1.0017(54) \\ 
  [1.0ex]
  \hline\\[-1.0ex]
6.3665 & 6 & 3.111(4) & 0.135488(6) & 1.0494(21) & 1.0317(31) & 0.9831(35) \\ 
6.6100 & 8 & 3.111(6) & 0.135339(3) & 1.0269(18) & 1.0191(44) & 0.9924(46) \\ 
6.9322 & 12 & 3.111(12) & 0.134855(3) & 1.0156(24) & 1.0200(55) & 1.0043(59) \\ 
7.1911 & 16 & 3.111(16) & 0.134411(3) & 0.9981(30) & 1.0006(49) & 1.0025(57) \\ 
  [1.0ex]
  \hline\\[-1.0ex]
6.2204 & 6 & 3.480(8) & 0.135470(15) & 1.0544(22) & 1.0390(33) & 0.9854(37) \\ 
6.4527 & 8 & 3.480(14) & 0.135543(9) & 1.0345(20) & 1.0330(68) & 0.9986(69) \\ 
6.7750 & 12 & 3.480(39) & 0.135121(5) & 1.0186(24) & 1.0279(46) & 1.0091(51) \\ 
7.0203 & 16 & 3.480(21) & 0.134707(4) & 1.0067(29) & 1.0233(58) & 1.0165(65) \\ 
  [1.0ex]
\hline\\[-1.0ex]
\end{tabular}
\caption{
Numerical values of the renormalization constant $\mathcal{Z'}_{2;0}^{+;(3)}$ 
and the step scaling function ${\Sigma}_{2;0}^{+;(3)}$ with HYP2 action at various
renormalized SF couplings and lattice spacings (continued).}
\label{tab:tab6}

}

\vfill\eject

\vbox{\ }
\vskip 0.48cm

\TABLE[!h]{
\footnotesize
\centering
\begin{tabular}{rrllccc}
\hline \\[-1.0ex]
$~~~~~~~\beta~~~~$ & ${L}/{a}$ & $~~~\gbar^2_{\rm\scriptscriptstyle SF}(L)$ &
$~~~~~\hopc$ & $\mathcal{Z'}_{3;0}^{+;(1)}\left(g_0,{a}/{L}\right)$ & $\mathcal{Z'}_{3;0}^{+;(1)}\left(g_0,{a}/{2L}\right)$ &\
 ${\Sigma}_{3;0}^{+;(1)}\left(g_0,{a}/{L}\right)$ \\[1.0ex]
\hline \\[-1.0ex]
10.7503 & 6 & 0.8873(5) & 0.130591(4) & 0.9377(5) & 0.9039(8) & 0.9640(10) \\ 
11.0000 & 8 & 0.8873(10) & 0.130439(3) & 0.9250(5) & 0.8913(20) & 0.9636(22) \\ 
11.3384 & 12 & 0.8873(30) & 0.130251(2) & 0.9093(5) & 0.8758(21) & 0.9632(24) \\ 
11.5736 & 16 & 0.8873(25) & 0.130125(2) & 0.8987(12) & 0.8578(36) & 0.9545(42) \\ 
  [1.0ex]
  \hline\\[-1.0ex]
10.0500 & 6 & 0.9944(7) & 0.131073(5) & 0.9313(6) & 0.8926(6) & 0.9584(9) \\ 
10.3000 & 8 & 0.9944(13) & 0.130889(3) & 0.9180(6) & 0.8785(23) & 0.9570(26) \\ 
10.6086 & 12 & 0.9944(30) & 0.130692(2) & 0.8993(5) & 0.8647(20) & 0.9615(23) \\ 
10.8910 & 16 & 0.9944(28) & 0.130515(2) & 0.8886(19) & 0.8504(30) & 0.9570(40) \\ 
  [1.0ex]
  \hline\\[-1.0ex]
9.5030 & 6 & 1.0989(8) & 0.131514(5) & 0.9258(7) & 0.8830(8) & 0.9538(11) \\ 
9.7500 & 8 & 1.0989(13) & 0.131312(3) & 0.9114(6) & 0.8693(23) & 0.9538(26) \\ 
10.0577 & 12 & 1.0989(40) & 0.131079(3) & 0.8921(8) & 0.8473(25) & 0.9498(29) \\ 
10.3419 & 16 & 1.0989(44) & 0.130876(2) & 0.8833(24) & 0.8364(28) & 0.9469(41) \\ 
  [1.0ex]
  \hline\\[-1.0ex]
8.8997 & 6 & 1.2430(13) & 0.132072(9) & 0.9182(7) & 0.8687(10) & 0.9461(13) \\ 
9.1544 & 8 & 1.2430(14) & 0.131838(4) & 0.9034(7) & 0.8523(25) & 0.9434(28) \\ 
9.5202 & 12 & 1.2430(35) & 0.131503(3) & 0.8825(7) & 0.8346(18) & 0.9457(22) \\ 
9.7350 & 16 & 1.2430(34) & 0.131335(3) & 0.8677(14) & 0.8180(29) & 0.9427(37) \\ 
  [1.0ex]
  \hline\\[-1.0ex]
8.6129 & 6 & 1.3293(12) & 0.132380(6) & 0.9145(8) & 0.8626(10) & 0.9432(13) \\ 
8.8500 & 8 & 1.3293(21) & 0.132140(5) & 0.8968(7) & 0.8435(33) & 0.9406(38) \\ 
9.1859 & 12 & 1.3293(60) & 0.131814(3) & 0.8763(9) & 0.8206(29) & 0.9364(35) \\ 
9.4381 & 16 & 1.3293(40) & 0.131589(2) & 0.8588(14) & 0.8093(26) & 0.9424(34) \\ 
  [1.0ex]
  \hline\\[-1.0ex]
8.3124 & 6 & 1.4300(20) & 0.132734(10) & 0.9093(8) & 0.8534(10) & 0.9385(13) \\ 
8.5598 & 8 & 1.4300(21) & 0.132453(5) & 0.8918(8) & 0.8405(31) & 0.9425(35) \\ 
8.9003 & 12 & 1.4300(50) & 0.132095(3) & 0.8702(7) & 0.8058(33) & 0.9260(39) \\ 
9.1415 & 16 & 1.4300(58) & 0.131855(3) & 0.8535(16) & 0.7962(26) & 0.9329(36) \\ 
  [1.0ex]
  \hline\\[-1.0ex]
7.9993 & 6 & 1.5553(15) & 0.133118(7) & 0.9035(9) & 0.8395(12) & 0.9292(16) \\ 
8.2500 & 8 & 1.5553(24) & 0.132821(5) & 0.8846(9) & 0.8284(34) & 0.9365(40) \\ 
8.5985 & 12 & 1.5533(70) & 0.132427(3) & 0.8600(16) & 0.8018(35) & 0.9323(44) \\ 
8.8323 & 16 & 1.5533(70) & 0.132169(3) & 0.8426(15) & 0.7830(27) & 0.9293(36) \\ 
  [1.0ex]
\hline\\[-1.0ex]
\end{tabular}
\caption{
Numerical values of the renormalization constant $\mathcal{Z'}_{3;0}^{+;(1)}$ 
and the step scaling function ${\Sigma}_{3;0}^{+;(1)}$ with HYP2 action at various
renormalized SF couplings and lattice spacings.}
\label{tab:tab7}

}

\vfill\eject

\vbox{\ }
\vskip 0.48cm

\TABLE[!h]{
\footnotesize
\centering
\begin{tabular}{rrllccc}
\hline \\[-1.0ex]
$~~~~~~~\beta~~~~$ & ${L}/{a}$ & $~~~\gbar^2_{\rm\scriptscriptstyle SF}(L)$ &
$~~~~~\hopc$ & $\mathcal{Z'}_{3;0}^{+;(1)}\left(g_0,{a}/{L}\right)$ & $\mathcal{Z'}_{3;0}^{+;(1)}\left(g_0,{a}/{2L}\right)$ &\
 ${\Sigma}_{3;0}^{+;(1)}\left(g_0,{a}/{L}\right)$ \\[1.0ex]
\hline \\[-1.0ex]
7.7170 & 6 & 1.6950(26) & 0.133517(8) & 0.8969(9) & 0.8296(9) & 0.9250(14) \\ 
7.9741 & 8 & 1.6950(28) & 0.133179(5) & 0.8773(7) & 0.8111(30) & 0.9245(35) \\ 
8.3218 & 12 & 1.6950(79) & 0.132756(4) & 0.8515(8) & 0.7881(33) & 0.9255(40) \\ 
8.5479 & 16 & 1.6950(90) & 0.132485(3) & 0.8356(16) & 0.7693(38) & 0.9207(49) \\ 
  [1.0ex]
  \hline\\[-1.0ex]
7.4082 & 6 & 1.8811(22) & 0.133961(8) & 0.8868(10) & 0.8210(20) & 0.9258(25) \\ 
7.6547 & 8 & 1.8811(28) & 0.133632(6) & 0.8674(9) & 0.7913(44) & 0.9123(51) \\ 
7.9993 & 12 & 1.8811(38) & 0.133159(4) & 0.8400(13) & 0.7706(30) & 0.9174(38) \\ 
8.2415 & 16 & 1.8811(99) & 0.132847(3) & 0.8192(25) & 0.7497(27) & 0.9152(43) \\ 
  [1.0ex]
  \hline\\[-1.0ex]
7.1214 & 6 & 2.1000(39) & 0.134423(9) & 0.8809(11) & 0.7993(15) & 0.9074(21) \\ 
7.3632 & 8 & 2.1000(45) & 0.134088(6) & 0.8594(10) & 0.7731(30) & 0.8996(37) \\ 
7.6985 & 12 & 2.1000(80) & 0.133599(4) & 0.8261(13) & 0.7430(32) & 0.8994(42) \\ 
7.9560 & 16 & 2.100(11) & 0.133229(3) & 0.8077(18) & 0.7293(32) & 0.9029(45) \\ 
  [1.0ex]
  \hline\\[-1.0ex]
6.7807 & 6 & 2.4484(37) & 0.134994(11) & 0.8664(18) & 0.7746(21) & 0.8940(31) \\ 
7.0197 & 8 & 2.4484(45) & 0.134639(7) & 0.8434(10) & 0.7454(27) & 0.8838(34) \\ 
7.3551 & 12 & 2.4484(80) & 0.134141(5) & 0.8097(16) & 0.7245(32) & 0.8948(44) \\ 
7.6101 & 16 & 2.448(17) & 0.133729(4) & 0.7931(22) & 0.7036(30) & 0.8872(45) \\ 
  [1.0ex]
  \hline\\[-1.0ex]
6.5512 & 6 & 2.770(7) & 0.135327(12) & 0.8572(13) & 0.7481(19) & 0.8727(26) \\ 
6.7860 & 8 & 2.770(7) & 0.135056(8) & 0.8303(11) & 0.7221(33) & 0.8697(41) \\ 
7.1190 & 12 & 2.770(11) & 0.134513(5) & 0.7979(12) & 0.6895(42) & 0.8641(55) \\ 
7.3686 & 16 & 2.770(14) & 0.134114(3) & 0.7753(22) & 0.6707(33) & 0.8651(49) \\ 
  [1.0ex]
  \hline\\[-1.0ex]
6.3665 & 6 & 3.111(4) & 0.135488(6) & 0.8488(13) & 0.7229(20) & 0.8517(27) \\ 
6.6100 & 8 & 3.111(6) & 0.135339(3) & 0.8182(12) & 0.6960(31) & 0.8506(40) \\ 
6.9322 & 12 & 3.111(12) & 0.134855(3) & 0.7832(17) & 0.6666(41) & 0.8511(56) \\ 
7.1911 & 16 & 3.111(16) & 0.134411(3) & 0.7593(22) & 0.6458(36) & 0.8505(53) \\ 
  [1.0ex]
  \hline\\[-1.0ex]
6.2204 & 6 & 3.480(8) & 0.135470(15) & 0.8384(14) & 0.6920(22) & 0.8254(29) \\ 
6.4527 & 8 & 3.480(14) & 0.135543(9) & 0.8074(13) & 0.6668(45) & 0.8259(57) \\ 
6.7750 & 12 & 3.480(39) & 0.135121(5) & 0.7715(16) & 0.6361(30) & 0.8245(43) \\ 
7.0203 & 16 & 3.480(21) & 0.134707(4) & 0.7445(21) & 0.6115(39) & 0.8214(57) \\ 
  [1.0ex]
\hline\\[-1.0ex]
\end{tabular}
\caption{
Numerical values of the renormalization constant $\mathcal{Z'}_{3;0}^{+;(1)}$ 
and the step scaling function ${\Sigma}_{3;0}^{+;(1)}$ with HYP2 action at various
renormalized SF couplings and lattice spacings (continued).}
\label{tab:tab8}
}

\vfill\eject

\vbox{\ }
\vskip 0.48cm

\TABLE[!h]{
\footnotesize
\centering
\begin{tabular}{rrllccc}
\hline \\[-1.0ex]
$~~~~~~~\beta~~~~$ & ${L}/{a}$ & $~~~\gbar^2_{\rm\scriptscriptstyle SF}(L)$ &
$~~~~~\hopc$ & $\mathcal{Z'}_{4;0}^{+;(1)}\left(g_0,{a}/{L}\right)$ & $\mathcal{Z'}_{4;0}^{+;(1)}\left(g_0,{a}/{2L}\right)$ &\
 ${\Sigma}_{4;0}^{+;(1)}\left(g_0,{a}/{L}\right)$ \\[1.0ex]
\hline \\[-1.0ex]
10.7503 & 6 & 0.8873(5) & 0.130591(4) & 0.9349(5) & 0.9189(7) & 0.9829(9) \\ 
11.0000 & 8 & 0.8873(10) & 0.130439(3) & 0.9290(4) & 0.9145(16) & 0.9844(18) \\ 
11.3384 & 12 & 0.8873(30) & 0.130251(2) & 0.9239(4) & 0.9074(15) & 0.9821(17) \\ 
11.5736 & 16 & 0.8873(25) & 0.130125(2) & 0.9199(9) & 0.9022(26) & 0.9808(29) \\ 
  [1.0ex]
  \hline\\[-1.0ex]
10.0500 & 6 & 0.9944(7) & 0.131073(5) & 0.9287(5) & 0.9099(5) & 0.9798(8) \\ 
10.3000 & 8 & 0.9944(13) & 0.130889(3) & 0.9228(5) & 0.9042(18) & 0.9798(20) \\ 
10.6086 & 12 & 0.9944(30) & 0.130692(2) & 0.9157(4) & 0.9001(16) & 0.9830(18) \\ 
10.8910 & 16 & 0.9944(28) & 0.130515(2) & 0.9132(16) & 0.8967(21) & 0.9819(28) \\ 
  [1.0ex]
  \hline\\[-1.0ex]
9.5030 & 6 & 1.0989(8) & 0.131514(5) & 0.9226(6) & 0.9016(7) & 0.9772(10) \\ 
9.7500 & 8 & 1.0989(13) & 0.131312(3) & 0.9167(5) & 0.8978(18) & 0.9794(21) \\ 
10.0577 & 12 & 1.0989(40) & 0.131079(3) & 0.9101(7) & 0.8911(17) & 0.9791(20) \\ 
10.3419 & 16 & 1.0989(44) & 0.130876(2) & 0.9071(18) & 0.8859(22) & 0.9766(32) \\ 
  [1.0ex]
  \hline\\[-1.0ex]
8.8997 & 6 & 1.2430(13) & 0.132072(9) & 0.9145(7) & 0.8897(8) & 0.9729(11) \\ 
9.1544 & 8 & 1.2430(14) & 0.131838(4) & 0.9092(6) & 0.8856(18) & 0.9740(20) \\ 
9.5202 & 12 & 1.2430(35) & 0.131503(3) & 0.9016(5) & 0.8800(15) & 0.9760(18) \\ 
9.7350 & 16 & 1.2430(34) & 0.131335(3) & 0.8972(12) & 0.8743(21) & 0.9745(26) \\ 
  [1.0ex]
  \hline\\[-1.0ex]
8.6129 & 6 & 1.3293(12) & 0.132380(6) & 0.9114(7) & 0.8855(8) & 0.9716(12) \\ 
8.8500 & 8 & 1.3293(21) & 0.132140(5) & 0.9038(6) & 0.8767(25) & 0.9700(28) \\ 
9.1859 & 12 & 1.3293(60) & 0.131814(3) & 0.8966(8) & 0.8718(20) & 0.9723(24) \\ 
9.4381 & 16 & 1.3293(40) & 0.131589(2) & 0.8906(11) & 0.8698(19) & 0.9766(24) \\ 
  [1.0ex]
  \hline\\[-1.0ex]
8.3124 & 6 & 1.4300(20) & 0.132734(10) & 0.9059(7) & 0.8775(8) & 0.9686(11) \\ 
8.5598 & 8 & 1.4300(21) & 0.132453(5) & 0.8991(7) & 0.8755(27) & 0.9738(31) \\ 
8.9003 & 12 & 1.4300(50) & 0.132095(3) & 0.8926(6) & 0.8606(23) & 0.9641(27) \\ 
9.1415 & 16 & 1.4300(58) & 0.131855(3) & 0.8853(12) & 0.8596(18) & 0.9710(25) \\ 
  [1.0ex]
  \hline\\[-1.0ex]
7.9993 & 6 & 1.5553(15) & 0.133118(7) & 0.9002(8) & 0.8658(10) & 0.9618(14) \\ 
8.2500 & 8 & 1.5553(24) & 0.132821(5) & 0.8929(8) & 0.8659(26) & 0.9698(31) \\ 
8.5985 & 12 & 1.5533(70) & 0.132427(3) & 0.8843(13) & 0.8582(27) & 0.9705(34) \\ 
8.8323 & 16 & 1.5533(70) & 0.132169(3) & 0.8786(12) & 0.8502(20) & 0.9677(26) \\ 
  [1.0ex]
\hline\\[-1.0ex]
\end{tabular}
\caption{
Numerical values of the renormalization constant $\mathcal{Z'}_{4;0}^{+;(1)}$ 
and the step scaling function ${\Sigma}_{4;0}^{+;(1)}$ with HYP2 action at various
renormalized SF couplings and lattice spacings.}
\label{tab:tab9}
}

\vfill\eject

\vbox{\ }
\vskip 0.48cm

\TABLE[!h]{
\footnotesize
\centering
\begin{tabular}{rrllccc}
\hline \\[-1.0ex]
$~~~~~~~\beta~~~~$ & ${L}/{a}$ & $~~~\gbar^2_{\rm\scriptscriptstyle SF}(L)$ &
$~~~~~\hopc$ & $\mathcal{Z'}_{4;0}^{+;(1)}\left(g_0,{a}/{L}\right)$ & $\mathcal{Z'}_{4;0}^{+;(1)}\left(g_0,{a}/{2L}\right)$ &\
 ${\Sigma}_{4;0}^{+;(1)}\left(g_0,{a}/{L}\right)$ \\[1.0ex]
\hline \\[-1.0ex]
7.7170 & 6 & 1.6950(26) & 0.133517(8) & 0.8933(8) & 0.8575(7) & 0.9599(12) \\ 
7.9741 & 8 & 1.6950(28) & 0.133179(5) & 0.8861(6) & 0.8541(23) & 0.9639(27) \\ 
8.3218 & 12 & 1.6950(79) & 0.132756(4) & 0.8784(7) & 0.8485(24) & 0.9660(28) \\ 
8.5479 & 16 & 1.6950(90) & 0.132485(3) & 0.8723(14) & 0.8398(28) & 0.9627(35) \\ 
  [1.0ex]
  \hline\\[-1.0ex]
7.4082 & 6 & 1.8811(22) & 0.133961(8) & 0.8833(9) & 0.8494(16) & 0.9616(21) \\ 
7.6547 & 8 & 1.8811(28) & 0.133632(6) & 0.8779(8) & 0.8392(30) & 0.9559(36) \\ 
7.9993 & 12 & 1.8811(38) & 0.133159(4) & 0.8682(10) & 0.8344(22) & 0.9611(27) \\ 
8.2415 & 16 & 1.8811(99) & 0.132847(3) & 0.8621(18) & 0.8267(20) & 0.9589(31) \\ 
  [1.0ex]
  \hline\\[-1.0ex]
7.1214 & 6 & 2.1000(39) & 0.134423(9) & 0.8768(10) & 0.8309(13) & 0.9477(18) \\ 
7.3632 & 8 & 2.1000(45) & 0.134088(6) & 0.8701(9) & 0.8245(23) & 0.9476(28) \\ 
7.6985 & 12 & 2.1000(80) & 0.133599(4) & 0.8568(10) & 0.8175(24) & 0.9541(30) \\ 
7.9560 & 16 & 2.100(11) & 0.133229(3) & 0.8526(14) & 0.8146(24) & 0.9554(32) \\ 
  [1.0ex]
  \hline\\[-1.0ex]
6.7807 & 6 & 2.4484(37) & 0.134994(11) & 0.8617(16) & 0.8114(17) & 0.9416(26) \\ 
7.0197 & 8 & 2.4484(45) & 0.134639(7) & 0.8554(9) & 0.8016(21) & 0.9371(26) \\ 
7.3551 & 12 & 2.4484(80) & 0.134141(5) & 0.8451(13) & 0.8032(24) & 0.9504(32) \\ 
7.6101 & 16 & 2.448(17) & 0.133729(4) & 0.8419(17) & 0.7942(23) & 0.9433(33) \\ 
  [1.0ex]
  \hline\\[-1.0ex]
6.5512 & 6 & 2.770(7) & 0.135327(12) & 0.8530(11) & 0.7872(16) & 0.9229(22) \\ 
6.7860 & 8 & 2.770(7) & 0.135056(8) & 0.8446(10) & 0.7814(27) & 0.9252(33) \\ 
7.1190 & 12 & 2.770(11) & 0.134513(5) & 0.8344(10) & 0.7730(29) & 0.9264(37) \\ 
7.3686 & 16 & 2.770(14) & 0.134114(3) & 0.8263(17) & 0.7686(26) & 0.9302(37) \\ 
  [1.0ex]
  \hline\\[-1.0ex]
6.3665 & 6 & 3.111(4) & 0.135488(6) & 0.8449(12) & 0.7654(16) & 0.9059(23) \\ 
6.6100 & 8 & 3.111(6) & 0.135339(3) & 0.8351(11) & 0.7604(25) & 0.9105(32) \\ 
6.9322 & 12 & 3.111(12) & 0.134855(3) & 0.8241(14) & 0.7551(30) & 0.9163(40) \\ 
7.1911 & 16 & 3.111(16) & 0.134411(3) & 0.8160(17) & 0.7501(27) & 0.9192(39) \\ 
  [1.0ex]
  \hline\\[-1.0ex]
6.2204 & 6 & 3.480(8) & 0.135470(15) & 0.8334(13) & 0.7384(17) & 0.8860(25) \\ 
6.4527 & 8 & 3.480(14) & 0.135543(9) & 0.8241(11) & 0.7351(37) & 0.8920(46) \\ 
6.7750 & 12 & 3.480(39) & 0.135121(5) & 0.8144(13) & 0.7319(24) & 0.8987(33) \\ 
7.0203 & 16 & 3.480(21) & 0.134707(4) & 0.8047(16) & 0.7239(30) & 0.8996(42) \\ 
  [1.0ex]
\hline\\[-1.0ex]
\end{tabular}
\caption{
Numerical values of the renormalization constant $\mathcal{Z'}_{4;0}^{+;(1)}$ 
and the step scaling function ${\Sigma}_{4;0}^{+;(1)}$ with HYP2 action at various
renormalized SF couplings and lattice spacings (continued).}
\label{tab:tab10}
}
\vfill\eject

\vbox{\ }
\vskip 0.48cm

\TABLE[!h]{
\centering
\begin{tabular}{l@{\hspace{10mm}}llll}
\hline\\[-1.0ex]
$~~~u$ &
$~{\sigma}_{1;0}^{+;(1)}$ &
$~{\sigma}_{2;0}^{+;(3)}$ &
$~{\sigma}_{3;0}^{+;(1)}$ &
$~{\sigma}_{4;0}^{+;(1)}$ \\[1.0ex]
\hline\\[-1.0ex]
0.8873 & 0.950(7) & 0.989(7) & 0.954(6) & 0.977(5) \\
0.9944 & 0.961(8) & 1.004(7) & 0.963(7) & 0.986(5) \\
1.0989 & 0.955(9) & 0.991(8) & 0.941(7) & 0.976(5) \\
1.2430 & 0.953(8) & 0.997(7) & 0.946(7) & 0.977(5) \\
1.3293 & 0.946(9) & 1.007(8) & 0.941(7) & 0.982(5) \\
1.4300 & 0.938(9) & 0.985(8) & 0.918(8) & 0.965(6) \\
1.5553 & 0.922(10) & 0.997(8) & 0.922(8) & 0.966(6)\\
1.6950 & 0.933(11) & 0.992(9) & 0.920(9) & 0.964(7) \\
1.8811 & 0.922(10) & 0.991(10) & 0.920(10) & 0.964(7) \\
2.1000 & 0.896(10) & 1.012(10) & 0.904(9) &  0.964(6) \\
2.4484 & 0.911(10) & 1.014(9) & 0.897(9) & 0.957(6) \\ 
2.770  & 0.862(11) & 1.008(11) & 0.859(10) & 0.934(8) \\ 
3.111  & 0.859(11) & 1.016(11) & 0.851(10) &  0.928(8)\\
3.480  & 0.817(12) & 1.033(14) & 0.818(12) & 0.908(9) \\
[1.0ex]\hline
\end{tabular}
\caption{
Continuum extrapolations of ${\Sigma}_{k;\alpha}^{+;(s)}$.
Linear dependence on $a/L$ is assumed. Data at $L/a=6$ have not been 
taken into account. 
}
\label{tab:CLe1}
}

\TABLE[!h]{
\centering
\begin{tabular}{rrl@{\hspace{10mm}}llll}
\hline \\[-1.0ex]
$\beta~~~$ & ${L}/{a}$ &
$~~~~~~\hopc$ & $~~\mathcal{Z'}_{1;0}^{+;(1)}$ & $~~\mathcal{Z'}_{2;0}^{+;(3)}$ & $~~\mathcal{Z'}_{3;0}^{+;(1)}$ & $~~\mathcal{Z'}_{4;0}^{+;(1)}$ \\[1.0ex]
\hline \\[-1.0ex]
6.0219 & 8  & 0.135043(17) & 0.7325(12)& 1.0749(19)& 0.7561(11)& 0.7795(10)\\ 
6.1628 & 10 & 0.135643(11) & 0.7058(12)& 1.0666(19)& 0.7263(12)& 0.7698(10)\\ 
6.2885 & 12 & 0.135739(13) & 0.6821(10)& 1.0518(16)& 0.7025(10)& 0.7598(8) \\ 
6.4956 & 16 & 0.135577(7)  & 0.6555(28)& 1.0414(44)& 0.6726(28)& 0.7499(22)\\ 
[1.0ex]
\hline
\end{tabular}
\caption{
Results for $\mathcal{Z'}_{k;\alpha}^{+;(s)}(g_0,a/L)$ at fixed scale $L=1.436 \,r_0$
(corresponding to $\mu_{\rm had}^{-1}=2L_{\rm max}$).
}
\label{tab:Zm1}
}
\vfill\eject

\vbox{\ }
\vskip 1.0cm

\vfill\eject

\vbox{\ }
\vskip 3.0cm

\vfill\eject

\vbox{\ }
\vskip 3.0cm

\vfill\eject

\vbox{\ }
\vskip 3.0cm

\vfill\eject

\vbox{\ }
\vskip 3.0cm

\vfill\eject

\vfill\eject

\vfill\eject


\clearpage

\begin{thebibliography}{99}

\bibitem{Okamoto:2005zg}
  M.~Okamoto,
  PoS {\bf LAT2005} (2006) 013
  [arXiv:hep-lat/0510113].

\bibitem{Onogi:2006km}
  T.~Onogi,
  PoS {\bf LAT2006} (2006) 017
  [arXiv:hep-lat/0610115].

\bibitem{Palombi:2006pu}
  F.~Palombi, M.~Papinutto, C.~Pena and H.~Wittig [ALPHA Collaboration],
  JHEP {\bf 0608} (2006) 017
  [arXiv:hep-lat/0604014].

\bibitem{Frezzotti:2000nk}
  R.~Frezzotti, P.A.~Grassi, S.~Sint and P.~Weisz  [ALPHA collaboration],
  JHEP {\bf 0108} (2001) 058
  [arXiv:hep-lat/0101001].

\bibitem{Donini:1999sf} A.~Donini, V.~Gim\'enez, G.~Martinelli,
  M.~Talevi and A.~Vladikas, 
  Eur.\ Phys.\ J.\ C {\bf 10} (1999) 121
  [arXiv:hep-lat/9902030].

\bibitem{Luscher:1992an}
M.~L\"uscher, R.~Narayanan, P.~Weisz and U.~Wolff,
Nucl.\ Phys.\ B {\bf 384} (1992) 168
[arXiv:hep-lat/9207009].

\bibitem{Sint:1993un}
  S.~Sint,
  Nucl.\ Phys.\ B {\bf 421}, 135 (1994)
  [arXiv:hep-lat/9312079].

\bibitem{alpha:su3}
M.~L\"uscher, R.~Sommer, P.~Weisz and U.~Wolff,
Nucl.\ Phys.\ B {\bf 413} (1994) 481
[arXiv:hep-lat/9309005].

\bibitem{DellaMorte:2004bc}
M.~Della Morte, R.~Frezzotti, J.~Heitger, J.~Rolf, R.~Sommer and
U.~Wolff [ALPHA Collaboration],
Nucl.\ Phys.\ B {\bf 713} (2005) 378
[arXiv:hep-lat/0411025].

\bibitem{mbar:pap1}
S.~Capitani, M.~L\"uscher, R.~Sommer and H.~Wittig  [ALPHA Collaboration],
Nucl.\ Phys.\ B {\bf 544} (1999) 669
[arXiv:hep-lat/9810063].

\bibitem{Guagnelli:2004za}
M.~Guagnelli, J.~Heitger, F.~Palombi, C.~Pena and A.~Vladikas
[ALPHA Collaboration],
JHEP {\bf 0405}, 001 (2004)
[arXiv:hep-lat/0402022].

\bibitem{DellaMorte:2005kg}
  M.~Della Morte, R.~Hoffmann, F.~Knechtli, J.~Rolf, R.~Sommer, I.~Wetzorke and U.~Wolff
  [ALPHA Collaboration],
  Nucl.\ Phys.\  B {\bf 729} (2005) 117
  [arXiv:hep-lat/0507035].

\bibitem{mbar:pap3}
J.~Garden, J.~Heitger, R.~Sommer and H.~Wittig  [ALPHA Collaboration],
Nucl.\ Phys.\ B {\bf 571} (2000) 237
[arXiv:hep-lat/9906013].

\bibitem{StF}
  M.~Guagnelli, K.~Jansen and R.~Petronzio,
  Phys.\ Lett.\  B {\bf 459} (1999) 594
  [arXiv:hep-lat/9903012];  
  \\[1.3ex]
  M.~Guagnelli, K.~Jansen, F.~Palombi, R.~Petronzio, A.~Shindler and I.~Wetzorke [Zeuthen-Rome / ZeRo Collaboration], 
  Nucl.\ Phys.\  B {\bf 664} (2003) 276
  [arXiv:hep-lat/0303012];  
  \\[1.3ex]
  M.~Guagnelli, K.~Jansen, F.~Palombi, R.~Petronzio, A.~Shindler and I.~Wetzorke [Zeuthen-Rome (ZeRo) Collaboration],
  Eur.\ Phys.\ J.\ C {\bf 40} (2005) 69
  [arXiv:hep-lat/0405027].

\bibitem{Heitger:2003xg}
  J.~Heitger, M.~Kurth and R.~Sommer  [ALPHA Collaboration],
  Nucl.\ Phys.\ B {\bf 669} (2003) 173
  [arXiv:hep-lat/0302019].

\bibitem{DellaMorte:2006sv}
  M.~Della Morte, P.~Fritzsch and J.~Heitger [ALPHA Collaboration],
  JHEP {\bf 0702}, 079 (2007)
  [arXiv:hep-lat/0611036].

\bibitem{Guagnelli:2005zc}
M.~Guagnelli, J.~Heitger, C.~Pena, S.~Sint and A.~Vladikas  [ALPHA
Collaboration],
JHEP {\bf 0603} (2006) 088
[arXiv:hep-lat/0505002].

\bibitem{Dimopoulos:2006dm}
  P.~Dimopoulos, J.~Heitger, F.~Palombi, C.~Pena, S.~Sint and A.~Vladikas [ALPHA Collaboration],
  Nucl.\ Phys.\ B {\bf 749} (2006) 69
  [arXiv:hep-ph/0601002].

\bibitem{Dimopoulos:2007cn}
  P.~Dimopoulos, J.~Heitger, F.~Palombi, C.~Pena, S.~Sint and A.~Vladikas [ALPHA Collaboration],
  [arXiv:hep-lat/0702017].

\bibitem{DellaMorte:2005yc}
  M.~Della Morte, A.~Shindler and R.~Sommer [ALPHA Collaboration],
  JHEP {\bf 0508} (2005) 051
  [arXiv:hep-lat/0506008].

\bibitem{Flynn:1990qz}
  J.M.~Flynn, O.F.~Hern\'andez and B.R.~Hill,
  Phys.\ Rev.\ D {\bf 43} (1991) 3709.

\bibitem{Gimenez:1998mw}
  V.~Gimenez,
  Nucl.\ Phys.\  B {\bf 401} (1993) 116;
  \\[1.3ex]
  V.~Gim\'enez and J.~Reyes,
  Nucl.\ Phys.\ B {\bf 545} (1999) 576 [arXiv:hep-lat/9806023]; 
  \\[1.3ex]
  J.~Reyes, ``C\'alculo de elementos de matriz d\'ebiles para hadrones
  $B$ con la HQET en el ret\'{\i}culo'', Ph.~D.~Thesis, University of
  Valencia, May 2001.


\bibitem{Necco:2001xg}
  S.~Necco and R.~Sommer,
  Nucl.\ Phys.\  B {\bf 622} (2002) 328 
  [arXiv:hep-lat/0108008].

\bibitem{Luscher:1996sc}
  M.~L\"uscher, S.~Sint, R.~Sommer and P.~Weisz,
  Nucl.\ Phys.\  B {\bf 478} (1996) 365
  [arXiv:hep-lat/9605038].

\bibitem{Sheikholeslami:1985ij}
  B.~Sheikholeslami and R.~Wohlert,
  Nucl.\ Phys.\  B {\bf 259}, 572 (1985).

\bibitem{Luscher:1996ug}
  M.~L\"uscher, S.~Sint, R.~Sommer, P.~Weisz and U.~Wolff,
  Nucl.\ Phys.\  B {\bf 491} (1997) 323
  [arXiv:hep-lat/9609035].

\bibitem{Bode:1998hd}
  A.~Bode, U.~Wolff and P.~Weisz  [ALPHA Collaboration],
   ``Two-loop computation of the Schroedinger functional in pure SU(3)  lattice
  Nucl.\ Phys.\ B {\bf 540} (1999) 491
  [arXiv:hep-lat/9809175].

\bibitem{Luscher:1996vw}
  M.~L\"uscher and P.~Weisz,
  Nucl.\ Phys.\  B {\bf 479}, 429 (1996)
  [arXiv:hep-lat/9606016].

\bibitem{Eichten:1989zv}
  E.~Eichten and B.~Hill,
  Phys.\ Lett.\ B {\bf 234} (1990) 511.

\bibitem{Hasenfratz:2001hp}
  A.~Hasenfratz and F.~Knechtli,
  Phys.\ Rev.\ D {\bf 64} (2001) 034504
  [arXiv:hep-lat/0103029].

\bibitem{Fischer:1996th}
  S.~Fischer, A.~Frommer, U.~Gl\"assner, T.~Lippert, G.~Ritzenh\"ofer and
  K.~Schilling,
  Comput.\ Phys.\ Commun.\  {\bf 98} (1996) 20
  [arXiv:hep-lat/9602019].

\bibitem{Guagnelli:1999nt}
  M.~Guagnelli and J.~Heitger  [ALPHA Collaboration],
  Comput.\ Phys.\ Commun.\  {\bf 130} (2000) 12
  [arXiv:hep-lat/9910024].

\bibitem{website}
{\tt http://www.kph.uni-mainz.de/T/lattice/ssf\_4quark\_static.pdf}

\bibitem{Palombi:2005zd}
  F.~Palombi, C.~Pena and S.~Sint [ALPHA Collaboration],
  JHEP {\bf 0603} (2006) 089
  [arXiv:hep-lat/0505003].

\bibitem{4fstatnf2}
  ALPHA Collaboration, in preparation

\end{thebibliography}
\end{document}